\documentclass{WileyMSP-template}
\usepackage{textcomp}
\usepackage{amsmath}
\usepackage{amssymb}
\usepackage{braket}
\usepackage{breqn}
\usepackage{amsthm}
\usepackage{appendix}
\usepackage{bbm}

\newtheorem{proposition}{Proposition}
\newtheorem{corollary}{Corollary}
\newtheorem{definition}{Definition}

\newcommand{\dd}{\,\mathrm{d}}
\newcommand{\tr}{\,\mathrm{tr}}
\newcommand{\e}{\,\mathrm{e}}
\newcommand{\E}{\,\mathbb{E}}

\begin{document}

\pagestyle{fancy}
%\rhead{\includegraphics[width=2.5cm]{vch-logo.png}}

\title{Variance of relative entropy over random states}

\maketitle

% Author: Please give full first and last names for authors and include * after the name of all corresponding authors

\author{Linfeng Wei}
and
\author{Lu Wei$^{*}$}
% \author{Author Three*}

% Dedication

%\dedication{Optional dedication here. If no dedication is required, please leave blank}

% Affiliations: Please provide adacemic titles (Prof. or Dr.) for all authors where applicable, and include an institutional email address for all corresponding authors
\begin{affiliations}
% Linfeng Wei and Lu Wei\\
Department of Computer Science, Texas Tech University, Lubbock, Texas 79409, USA\\
Email Address: \{linwei, luwei\}@ttu.edu\\
$^{*}$Corresponding author
\end{affiliations}


% Abstract should be written in the present tense and impersonal style (i.e., avoid we), and be at most 200 words long
\begin{abstract}
Average relative entropy over major generic random states is recently studied. In this work, the investigation is extended by deriving the variances of relative entropy over the Hilbert–Schmidt and Bures–Hall ensembles, which characterize its fluctuations around the average value. Exact results for the variances of relative entropy between independent states drawn from the same ensemble and from two different ensembles are obtained. These results further indicate that the average relative entropy is typical in the considered limiting regimes. In addition to the matrix-level calculations similar to the average relative entropy, eigenvalue correlation functions of the ensembles are involved in the variance computations through the use of zonal polynomials.
\end{abstract}

% Text: Please use section headings and subheadings as specified below. For communications, all section headings apart from Experimental Section should be removed
% Please make the first reference to a display item bold: \textbf{Figure 1}
% Do not abbreviate Figure, Equation, etc.; display items are always singular, i.e., Figure 1 and 2.
% Equations are always singular, i.e., Equation 1 and 2, and should be inserted using the {equation} environment, not as graphics
% Please do not use footnotes in the text, additional information can be added to the Reference list.

\section{Introduction}
Random states play an important role in modern quantum science, with applications ranging from quantum circuit complexity~\cite{Brandao}, quantum device benchmarking~\cite{Choi} to entanglement estimation~\cite{Page93}. The statistical properties of entropic quantities, such as the entanglement entropy and capacity, have been extensively investigated for individual random state drawn from a generic ensemble, including two of the most prominent ones as the Hilbert–Schmidt~(HS) ensemble~\cite{Page93,Foong94,Ruiz95,VPO16,Wei17,Wei20,HWC21} and the Bures–Hall~(BH) ensemble~\cite{Sarkar19,Wei20BHA,Wei20BH,wei2025skewnessvonneumannentropy}. In contrast, considerably less attention has been devoted to the study of distinguishability measures between pairs of random states, whether sampled from the same ensemble or from different ensembles, in which case the most fundamental metric is the relative entropy~\cite{Leditzky16,Kudler21,KudlerNR,wei2025averagerelativeentropyrandom}.

Relative entropy is a fundamental quantity in quantum information theory. It quantifies the distinguishability between two quantum states and measures the information loss incurred when one state is used to approximate another. Its classical counterpart is the Kullback–Leibler (KL) divergence, which characterizes the discrepancy between two probability distributions~\cite{Chen25}. In quantum information theory, relative entropy extends the KL divergence to density operators, thereby providing a natural measure of statistical distinguishability in quantum systems. Relative entropy possesses several important mathematical properties, including non-negativity, joint convexity, and monotonicity under completely positive trace-preserving maps~\cite{Leditzky16}. These properties guarantee that the distinguishability between quantum states cannot increase under physically realizable operations, making relative entropy a robust and operationally meaningful quantity. Relative entropy also provides a quantitative framework for studying deviations from equilibrium, assessing correlations, and analyzing information-theoretic aspects of both finite-dimensional and high-dimensional quantum systems~\cite{wei2025averagerelativeentropyrandom}.

A density matrix $\rho$ is a positive semidefinite matrix in a Hilbert space satisfying
\begin{equation}\label{eq:dm}
\tr\rho=1.
\end{equation}
The relative entropy between two density matrices $\rho$ and $\sigma$ of the same dimension $m$ is defined as~\cite{Leditzky16,Kudler21,wei2025averagerelativeentropyrandom},
\begin{equation}\label{eq:red}
D(\rho||\sigma)=\tr\left(\rho\left(\ln\rho-\ln\sigma\right)\right).
\end{equation}
Relative entropy~(\ref{eq:red}) enjoys various information-theoretic properties~\cite{Leditzky16}, including non-negativity
\begin{equation}
D(\rho||\sigma)\geq0,
\end{equation}
where equality $D(\rho||\sigma)=0$ holds for identical density matrices $\rho=\sigma$, and monotonicity
\begin{equation}
D(\rho||\sigma)\geq D\left(\Phi(\rho)||\Phi(\sigma)\right),
\end{equation}
where $\Phi$ is any completely positive trace-preserving map.

For two density matrices $\rho$ and $\sigma$ from random states, the relative entropy~(\ref{eq:red}) between them clearly becomes a random variable~\cite{wei2025averagerelativeentropyrandom}. As discussed in~\cite{Kudler21,KudlerNR}, closed-form expressions for the relative entropy of random quantum states are valuable across multiple areas of quantum information theory, including quantum hypothesis testing~\cite{Hiai91,Sreekumar25} and the eigenstate thermalization hypothesis~\cite{He17}. In particular,~\cite{Hiai91} showed that relative entropy provides a proper information-theoretic framework for quantum hypothesis testing and the asymptotic behavior of many identically prepared systems underpinning the thermodynamics. In the literature, an approximate expression of average relative entropy of random states from HS was obtained in~\cite{Kudler21}. Exact formulas of average relative entropy of random states from HS ensemble, BH ensemble, and between the random states from such two ensembles have been recently derived in~\cite{wei2025averagerelativeentropyrandom}, which we recall by listing them in~(\ref{avresult1})--(\ref{avresult3}). These results provided a first step to the study of statistical properties of relative entropy over the generic random state models of HS and BH ensembles, which are the most important random state models of bipartite systems, whose physical relevance is discussed in Section~\ref{appd:ensembles}.

In this work, we extend the investigation to the second order statistics by deriving exact formulas for the variance of the relative entropy over random states from such ensembles, respectively, as well as for the case where two random state $\rho$ and $\sigma$ are from different ensembles. The study of variance of relative entropy serves as an important step towards the characterization of its statistical distribution by describing the fluctuation around the average. This second-order information of the distribution of the relative entropy over random states is crucial, especially in quantum-informatics perspectives. Firstly, at finite dimensions, which is what real systems have, the mean can be misleading if the variance is large~\cite{Page93,Wei17}. A small variance means most instances cluster tightly around the mean, so the average is genuinely representative. A large variance means the system is highly dispersed, and the average may describe something that is actually rare. Secondly, in applications including quantum hypothesis testing, variance is directly tied to performance guarantees~\cite{Sreekumar25}. Two states might have a well-defined average distinguishability, but what matters operationally is how often discrimination succeeds or fails. Large fluctuations can lead to occasional hard-to-distinguish cases even if the average looks favorable. Thirdly, from a geometric viewpoint, variance indicates the “spread” of the information geometry of states~\cite{BZ06}. If relative entropy has low variance across an ensemble, then the ensemble is tightly structured; if it has high variance, the geometry is more irregular and state-dependent. Moreover, the asymptotic statistical behavior of the variance also enables the investigation on typicality that describes the tendency of a quantity to remain close to its average in large systems, allows us to predict the behavior of the relative entropy in most instances, despite the underlying randomness~\cite{BHK21,BHKRV22,Wei22}.

The variance of relative entropy consists of more terms that are not simple combinations of results on entanglement entropy cumulants. From the definition~(\ref{eq:red}), the variance of relative entropy between two density matrices $\rho$ and $\sigma$ is given by
\begin{eqnarray}
 \mathbb{V}[D(\rho \| \sigma)]  & =& \E\!\left[D^2(\rho \| \sigma)\right] -\E^2\!\left[D(\rho \| \sigma)\right]\label{eq:rev}\\
 &=&\mathbb{V}[\operatorname{tr}(\rho \ln \rho)]+2 \mathbb{E}[\operatorname{tr}(\rho \ln \rho)] \mathbb{E}[\operatorname{tr}(\rho \ln \sigma)] \nonumber\\&&+\mathbb{V}[\operatorname{tr}(\rho \ln \sigma)]-2 \mathbb{E}[\operatorname{tr}(\rho \ln \rho) \operatorname{tr}(\rho \ln \sigma)],\label{eq:rev4}
 \end{eqnarray}
 where the terms $\mathbb{E}[\operatorname{tr}(\rho \ln \rho)]$ and $\mathbb{V}[\operatorname{tr}(\rho \ln \rho)]$ are, up to a negative sign, the average and variance of entanglement entropy of reduced density matrix $\rho$ of a bipartite system. In the literature, exact formulas for the two terms have been obtained for HS ensemble in~\cite{Page93,Foong94,Ruiz95} and~\cite{VPO16,Wei17}, respectively,
 \begin{eqnarray}
   \mathbb{E}\left[-\operatorname{tr}\!\left(\rho_{\mathrm{HS}}\ln \rho_{\mathrm{HS}}\right)\right]&=& \psi_0(m n+1)-\psi_0(n)-\frac{m+1}{2 n} \label{EHS}\\
     \mathbb{V}\left[-\operatorname{tr}\!\left(\rho_{\mathrm{HS}}\ln \rho_{\mathrm{HS}}\right)\right]&=&-\psi_1(m n+1)+\frac{m+n}{m n+1} \psi_1(n)-\frac{(m+1)(m+2 n+1)}{4 n^2(m n+1)}, \label{VHS}
\end{eqnarray}
as well as for BH ensemble in~\cite{Sarkar19,Wei20BHA} and~\cite{Wei20BH}
      \begin{eqnarray}
       \mathbb{E}\left[-\operatorname{tr}\!\left(\rho_{\mathrm{BH}}\ln \rho_{\mathrm{BH}}\right)\right]&=&\psi_0\!\left(m n-\frac{m^2}{2}+1\right)-\psi_0\!\left(n+\frac{1}{2}\right) \label{EBH}\\
      \mathbb{V}\left[-\operatorname{tr}\!\left(\rho_{\mathrm{BH}}\ln \rho_{\mathrm{BH}}\right)\right]&=&-\psi_1\!\left(m n-\frac{m^2}{2}+1\right)+\frac{2 n(2 n+m)-m^2+1}{2 n\left(2 m n-m^2+2\right)} \psi_1\!\left(n+\frac{1}{2}\right),\label{VBH}
\end{eqnarray}
where the dimension of density matrices $\rho_{\rm{HS}}$ and $\rho_{\rm{BH}}$ is $m$ with parameter $n$, cf.~(\ref{eq:a}). The function $\psi_{k}(z)$ denotes the $k$-th order polygamma function~\cite{Prudnikov86}
\begin{equation}\label{eq:polygamma}
\psi_{k}(z)=\frac{\dd^{k+1}}{\dd z^{k+1}}\ln\Gamma(z)=\frac{\dd^{k}}{\dd z^{k}}\psi_{0}(z).
\end{equation}
The required terms for calculating variance of relative entropy then boils down to two major tasks on $\mathbb{E}[\operatorname{tr}(\rho \ln \rho) \operatorname{tr}(\rho \ln \sigma)]$ and $\mathbb{V}[\operatorname{tr}(\rho \ln \sigma)]$ in~(\ref{eq:rev4}), which are treated with the ingredients of zonal polynomials in Section~\ref{sec:zon}.

As a significant application of the variance of relative entropy, we also study the typicality of such distinguishability metric of random states in this work. In general, we say that an entropic metric, such as an entanglement entropy or relative entropy, is typical if its variance is insignificant in value~\cite{BHK21,BHKRV22,Wei22}. More precisely, one can conclude the typicality if, under certain limiting regime, the average of the metric is a constant up to some scaling in dimension, and the variance vanishes with the same scaling as the dimension increases. The average of entanglement entropy over different generic state models has been concluded to be typical for relatively large dimensions~\cite{Page93,VPO16,BHK21,BHKRV22}. In this work, we will utilize the exact formulas of variance of relative entropy derived for different state models to deduce the typicality by studying the limiting behavior of variance formulas under certain regime~(\ref{eq:limfixn-m}).

The rest of the paper is organized as follows. In Section~\ref{sec:results}, we state the exact formulas on variance of relative entropy of HS ensemble, BH ensemble and between the two ensembles, respectively, and then study their asymptotic behavior and typicality. In Section~\ref{sec:calc}, we first define the two ensembles before the detailed derivation of the variance formulas by applying the theory of zonal polynomials, and then perform the proof of the propositions and corollary from Section~\ref{sec:results}.

\section{Main results on variances of relative entropy}\label{sec:results}
The main results on exact variance of relative entropy are summarized as three propositions below. Despite the exact results of variance of relative entropy over entanglement entropy are obtained as purely closed-form expressions for HS~(\ref{eq:HS}) and BH ensemble~(\ref{eq:BH}), the results of variance of relative entropy are presented in terms of polygamma functions and finite summations of rational functions, while in a degenerate case when $n_1=n_2=m$, the results reduce to closed-form expressions. We first state the result on variance of relative entropy over HS ensemble~(\ref{eq:HS}).
\begin{proposition}\label{prop:HS}
For density matrices $\rho_{\rm{HS}}$ and $\sigma_{\rm{HS}}$ of HS ensemble~(\ref{eq:HS}) of system dimension $m$ with parameters of environment dimension $n_1$ and $n_2$, respectively, the variance of relative entropy~(\ref{eq:rev}) is given by
\begin{eqnarray}\label{eq:r1}
\mathbb{V}\!\left[D(\rho_{\rm{HS}}||\sigma_{\rm{HS}})\right]&=&\frac{2}{m n_1+1}\Psi_{0,\alpha_2}+\rm{CF_1}.
\end{eqnarray}
For a degenerate case when $n_1=n_2=m$,~(\ref{eq:r1}) reduces to
\begin{eqnarray}\label{eq:degenr1}
&&\mathbb{V}\!\left[D(\rho_{\rm{HS}}||\sigma_{\rm{HS}})\right]\nonumber\\
&=&\frac{m^4-4 m^3-m^2+4}{4 m^4 \left(m^2+1\right)}+\frac{1 }{m^2+1}\left(\frac{1}{m}(m-2)(\psi_0 (m)-\psi_0 (1))+(3 m-1) \psi _1(m)+2 \psi _1(1)\right)\nonumber\\&&-2 \psi _1\!\left(m^2\right).
    \end{eqnarray}
\end{proposition}
The proof of Proposition~\ref{prop:HS} is in Section~\ref{sec:prhs}. Here, the $\Psi_{0,\alpha_2}$ appeared in variance result is a function of finite summation with parameter
\begin{eqnarray}\label{eq:alphas}
\alpha_{i} = n_i-m, ~~i=1,2,
\end{eqnarray}
defined from the parameters of the given ensembles, where we denote a set of finite summations for finite $a$ and $b$, whenever exists, as
\begin{equation}\label{Psi}
\Psi_{a, b}=\frac{(m+a)!}{(m+b)!} \sum_{k=1}^{m+a} \frac{(m+b-k)!}{(m+a-k)!} \frac{1}{k^2} .
\end{equation}
Also, for simplicity, we use $\rm{CF_1}$ in~(\ref{eq:r1}) to represent closed-form expressions involving polygamma functions in the formulas of main results, where the complete expression is presented in the Appendix~\ref{appd:closedform}, and similarly for $\rm{CF_2}$, $\rm{CF_3}$, $\rm{CF_4}$ in~(\ref{eq:r2}), (\ref{eq:r3}), (\ref{eq:r4}), respectively.

We then proceed to present the result on variance of relative entropy BH ensemble~(\ref{eq:BH}).
\begin{proposition}\label{prop:BH}
For density matrices $\rho_{\rm{BH}}$ and $\sigma_{\rm{BH}}$ of BH ensemble~(\ref{eq:BH}) of system dimension $m$ with parameters of environment dimension $n_1$ and $n_2$, respectively, the variance of relative entropy~(\ref{eq:rev}) is given by
\begin{eqnarray}\label{eq:r2}
\mathbb{V}\!\left[D(\rho_{\rm{BH}}||\sigma_{\rm{BH}})\right]
=\frac{m-4 n_1}{n_1 \left(m^2-2 m n_1-2\right)}\mathcal{B_{\rm{BH}}} +\rm{CF_2},
\end{eqnarray}
For a degenerate case when $n_1=n_2=m$,~(\ref{eq:r2}) reduces to
\begin{eqnarray}\label{eq:degenr2}
&&\mathbb{V}\!\left[D(\rho_{\rm{BH}}||\sigma_{\rm{BH}})\right]\nonumber\\
&=&\frac{3 m^4+5 m^2+16}{2 m^4 \left(m^2+2\right)}+\frac{1}{2 \left(m^2+2\right)}\bigg(\frac{3 }{m^2}(m-1) (2 m-1)(\psi_0 (m)-\psi_0 (1))+15 \psi _1(1)\nonumber\\&&+\frac{1}{m}\left(3 \left(2 m^2-2 m+1\right)\psi _1(m)+\left(5 m^2+1\right) \psi _1\!\left(m+\frac{1}{2}\right)\!\right)\!\!\bigg)-2 \psi _1\!\left(\frac{m^2}{2}\right).
\end{eqnarray}
\end{proposition}
The proof of Proposition~\ref{prop:BH} is in Section~\ref{sec:prbh}. Here, the function of Bures-Hall finite summation $ \mathcal{B_{\rm{BH}}}$  denotes \begin{eqnarray}\label{eq:bbh}
  \mathcal{B_{\rm{BH}}}= \Psi_{0,\beta_2}-\Psi_{2 \beta_2,\beta_2}+2 \Psi_{0,2 \beta_2}-2 \Psi_{0,2 \beta_2+m}.
\end{eqnarray}
with $\Psi_{a, b}$ in~(\ref{Psi}), and \begin{eqnarray}\label{eq:beta}
~~\beta_{i} = n_i-m-\frac{1}{2}, ~~i=1,2.
\end{eqnarray}
defined from the parameters of the given ensembles.

 Besides the relative entropy of random states from a certain ensemble, we also consider the case that two density matrices are from different ensembles, which is only possible when considering relative entropy rather than the well-studied entanglement entropy over a single ensemble. Since BH ensemble is a natural generalization model of the HS ensemble~\cite{wei2025averagerelativeentropyrandom}, we consider the case that $\rho$ is from BH ensemble~(\ref{eq:BH}) and $\sigma$ is from HS ensemble~(\ref{eq:HS}), which is of the information-theoretic interest as $\rho$ may be realized as full probabilistic model while $\sigma$ could be a less informative reduced-order approximate model for accelerate computations~\cite{Chen25,wei2025averagerelativeentropyrandom}.
\begin{proposition}\label{prop:HS-BH}
For density matrices $\rho_{\rm{BH}}$ of BH ensemble~(\ref{eq:BH}) and $\sigma_{\rm{HS}}$ of HS ensemble~(\ref{eq:HS}) of system dimension $m$ with parameters of environment dimension $n_1$ and $n_2$, respectively, the variance of relative entropy~(\ref{eq:rev}) is given by
\begin{eqnarray}\label{eq:r3}
\mathbb{V}\!\left[D(\rho_{\rm{BH}}||\sigma_{\rm{HS}})\right]&=&\frac{m-4 n_1}{n_1 \left(m^2-2 m n_1-2\right)}\Psi_{0,\alpha_2}+\rm{CF_3}.
\end{eqnarray}
For a degenerate case when $n_1=n_2=m$,~(\ref{eq:r3}) reduces to
\begin{eqnarray}\label{eq:r3n1n2=m}
&&\mathbb{V}\!\left[D(\rho_{\rm{BH}}||\sigma_{\rm{HS}})\right]\nonumber\\&=& \frac{3 m^4+5 m^2+16}{2 m^4(m^2+2) }+\frac{1}{m^2+2}\left(\frac{1}{2 m^2 }\left(2 m^2-6 m+1\right) (\psi_0 (m)-\psi_0 (1))+3 \psi _1(1)+\frac{1}{2m}\right.\nonumber\\&& \left.\times\left(\left(2 m^2-3 m+1\right)\psi _1(m)+\left(5 m^2+1\right) \psi_1\!\left(m+\frac{1}{2}\right)\!\right)\!\right)-\psi _1\!\left(\frac{m^2}{2}\right)-\psi_1\!\left(m^2\right).
\end{eqnarray}
\end{proposition}
The proof of Proposition~\ref{prop:HS-BH} is in Section~\ref{sec:prhsbh}. Note that an exact formula of $\mathbb{V}\!\left[D(\rho_{\rm{HS}}||\sigma_{\rm{BH}})\right]$ can be similarly derived, which, despite being less relevant from an information-theoretic perspective, is presented in~(\ref{eq:r4}) for completeness.

The closed-form results in the variance of relative entropy are expressed in terms of polygamma functions of positive integer and positive half-integer arguments, which also admit finite-sum representations as below. For positive integer argument $l$,
\begin{subequations}
\begin{eqnarray}
\psi_0(l)&=&-\gamma+\sum_{\smash{i=1}}^{l-1}\frac{1}{i}\\
\psi_0\!\left(l+\frac{1}{2}\right)&=&-\gamma-2 \ln 2+2 \sum_{i=0}^{l-1} \frac{1}{2 i+1}\\
\psi_1\left(l\right)&=&\zeta(2)- \sum_{i=1}^{l-1} \frac{1}{i^2}\\
\psi_1\!\left(l+\frac{1}{2}\right)&=&3\zeta(2)-4 \sum_{i=0}^{l-1} \frac{1}{(2 i+1)^2}
\end{eqnarray}
\end{subequations}
where $\gamma \approx 0.5772$ is the Euler's constant and
\begin{eqnarray}\label{zeta}
    \zeta(s)=\sum_{\smash{i=1}}^{\infty}\frac{1}{i^s}
\end{eqnarray} is the Riemann zeta function.

Before diving into the calculation of the exact results, we first state some asymptotic results as an application of the exact formulas~(\ref{eq:r1}), (\ref{eq:r2}),~(\ref{eq:r3}), which also allow us to study the typical behavior of relative entropy~(\ref{eq:red}). The typicality of an entropic metric is defined as follows.
\begin{definition}
    Suppose the limiting average of a metric function $S$ over random density matrix $X_n$ of dimension $n$ is a constant up to a scale of some order of dimension when $n \to \infty$, i.e.
\begin{eqnarray}\label{typicality1}
    \E\left[\frac{S(X_n)}{n^k}\right] \to c, ~~ n \to \infty, ~~ \textrm{for some constant } c, k,
\end{eqnarray}
then the average $\E\left[S(X_n)\right] $ is typical if
\begin{eqnarray}\label{typicality2}
    \mathbb{V}\left[\frac{S(X_n)}{n^k}\right] \to 0, ~~ n \to \infty.
\end{eqnarray}
\end{definition}
In particular, relative entropy exhibits typicality if its mean is typical under a certain regime. We now consider the asymptotic behavior of the variance of relative entropy. Throughout the asymptotic analysis, we consider a regime where the parameters $n_1$, $n_2$ and $m$ goes to infinity, with the assumption denoted as $F$ that the differences between each pairs of parameters $n_1$ and $m$, $n_2$ and $m$ are fixed, respectively.
\begin{corollary}\label{vanish}[Corollary of Propositions~\ref{prop:HS}--\ref{prop:HS-BH}]
In the asymptotic regime:
\begin{equation}\label{eq:limfixn-m}
m\to\infty, ~~n_i\to \infty,~~ i=1,2,~~ \text{with assumption F,}
\end{equation}
where $F$ denotes assumption:
\begin{eqnarray}\label{eq:assum}
    n_1-m,~~ n_2-m ~~\text{are both fixed constants},
\end{eqnarray}
the variances of relative entropy~(\ref{eq:r1}),~(\ref{eq:r2}),~(\ref{eq:r3}) of HS ensemble~(\ref{eq:HS}), BH ensemble~(\ref{eq:BH}), and between these two ensembles vanish,
\begin{eqnarray}\label{eq:vanish}
    \mathbb{V}\!\left[D(\rho||\sigma)\right]\longrightarrow 0, \qquad m\to \infty
\end{eqnarray}
for $\rho$ and $\sigma$ in~(\ref{eq:r1}),~(\ref{eq:r2}),~(\ref{eq:r3}). Moreover, the relative entropies of HS ensemble~(\ref{eq:HS}), BH ensemble~(\ref{eq:BH}), and between these two ensembles are typical under asymptotic regime~(\ref{eq:limfixn-m}).
\end{corollary}
The proof of the vanishing result of variance~(\ref{eq:vanish}) is straightforward from the asymptotics behavior of polygamma and finite summation functions under the regime~(\ref{eq:limfixn-m}), and such result along with the limiting behavior of the average of relative entropy obtained in~\cite{wei2025averagerelativeentropyrandom} gives the proof of the typicality according to its definition in~(\ref{typicality1}), (\ref{typicality2}). Both of the proofs are performed in Section~\ref{sec:prcoro}.

\begin{figure}[htbp]
  \centering
  \includegraphics[width=0.8\textwidth]{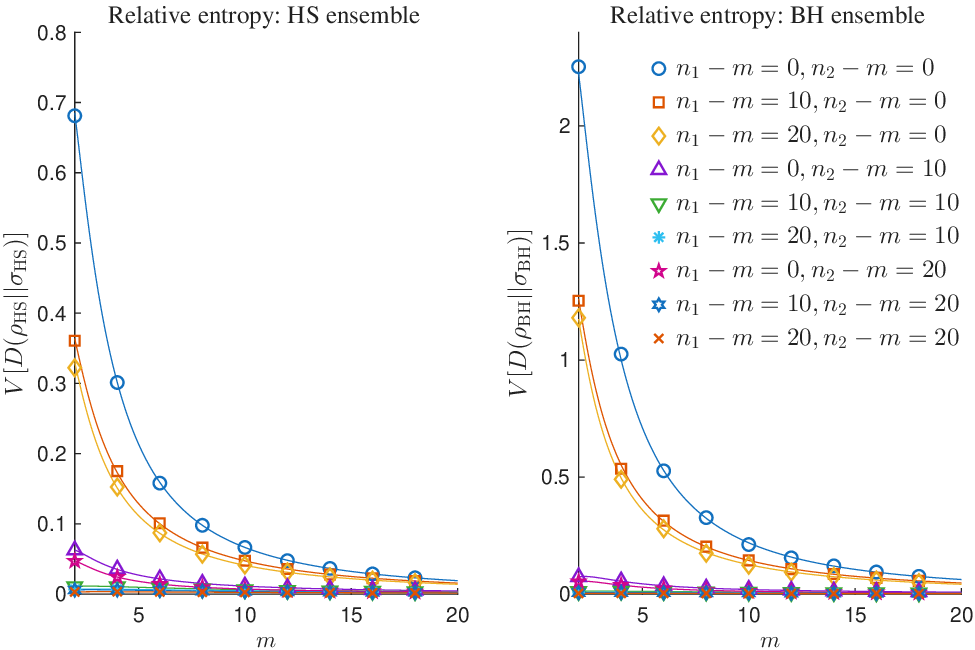}
  % where an .eps filename suffix will be assumed under latex,
  % and a .pdf suffix will be assumed for pdflatex
  \caption{Variance of relative entropy of states from the same ensemble: analytical results versus simulations. Solid lines are drawn by the exact formulas~(\ref{eq:r1}),~(\ref{eq:r2}). The symbols of circle, square, diamond, upward-pointing triangle, downward-pointing triangle, asterisk, pentagram, hexagram and cross, represent numerical simulations.}
  \label{fig:1}
\end{figure}

\begin{figure}[htbp]
  \centering
  \includegraphics[width=0.8\textwidth]{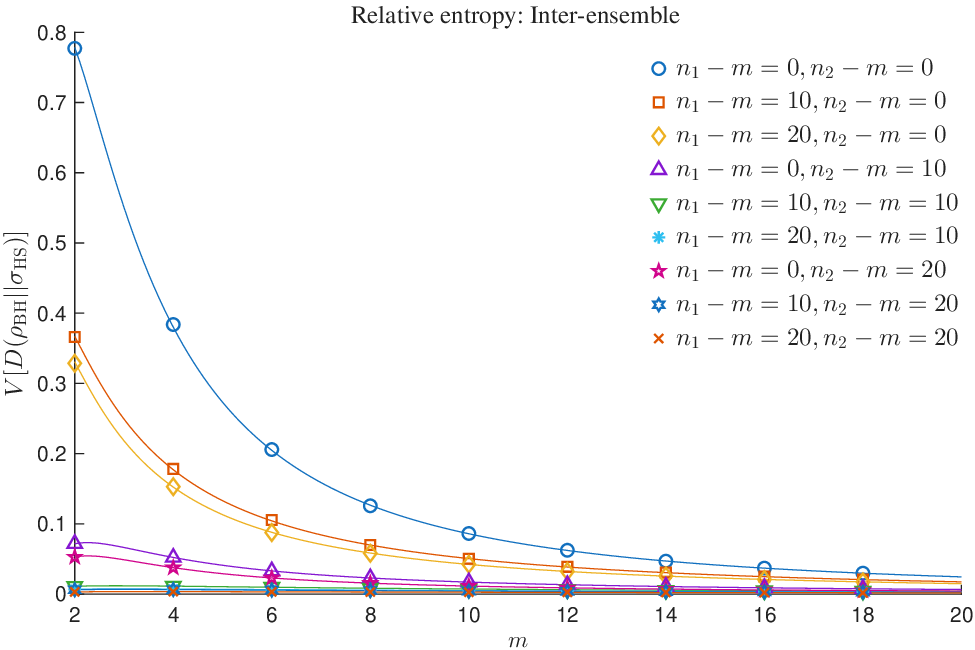}
  % where an .eps filename suffix will be assumed under latex,
  % and a .pdf suffix will be assumed for pdflatex
  \caption{Variance of relative entropy of states from two different ensembles: analytical results versus simulations. Solid lines are drawn by the exact formulas~(\ref{eq:r3}). The symbols of circle, square, diamond, upward-pointing triangle, downward-pointing triangle, asterisk, pentagram, hexagram and cross, represent numerical simulations.}
  \label{fig:2}
\end{figure}

Before going to the derivations in Section~\ref{sec:calc}, we conduct several numerical studies which are summarized in the two figures. In Figure~\ref{fig:1}, we plot variances of relative entropy of random states from the same ensemble, with the left-hand side and right-hand
side subfigures corresponding to the HS and BH ensembles, respectively. In Figure~\ref{fig:2}, we plot variances of relative entropy of random states from two different ensembles, corresponding to the case in Proposition~\ref{prop:HS-BH}. For each ensemble, we consider three different values $0, 10, 20$ for each of the two parameters $n_1-m$ and $n_2-m$, respectively. The numerical simulations, illustrated by the symbols with each variance taken over $10^5$
realizations of density matrices, match well with exact formulas~(\ref{eq:r1}),~(\ref{eq:r2}),~(\ref{eq:r3}). The simulation is performed by random sampling Wishart matrix for Hilbert--Schmidt ensemble~\cite{Page93,Ruiz95,huang2025cumulantstructuresentanglemententropy}, and sampling random unitary matrices from Bures--Hall measure by implementing the Metropolis–Hastings algorithm~\cite{Metropolis,Hastings}, cf.~\cite{Sarkar19} with energy function
\begin{equation}
    E(U) = -2\alpha  \ln \left|\det(I_m+U)\right|
\end{equation} for Bures--Hall ensemble, respectively. From the figures, one can also observe a tendency for the variances of the relative entropy~(\ref{eq:r1}),~(\ref{eq:r2}),~(\ref{eq:r3}) to decrease toward zero as the system dimension $m$ increases for selected instances of the regime~(\ref{eq:limfixn-m}). Nevertheless, one can also observe that, for fixed dimension $m$, an increase in either $n_1$ or $n_2$ would reduce the fluctuation in all the three cases, where an increase in $n_2$ reduces more fluctuation than the same increase of $n_1$ with the other fixed. As for a comparison between different cases, one can observe the HS ensemble relative entropy fluctuations are relatively less than BH ensemble ones, and the inter-ensemble behaves similar to the HS ensemble with a slightly larger fluctuation at same $m$ and fixed parameters.

There are certain interpretations and implications of the exact variance formulas and asymptotic behaviors above. Firstly, the exact formulas of variance of relative entropy~(\ref{eq:r1}), (\ref{eq:r2}), (\ref{eq:r3}) consist of finite summations of rational functions and polygamma functions in terms of system dimension $m$ and parameters of environment dimension $n_1$ from random state $\rho$, and $n_2$ from random state $\sigma$, where the parameter $n_2$ affects the variance value more significantly than $n_1$. This suggests that in quantum information processing and quantum hypothesis testing, increasing the environment dimension $n_2$ of random state model $\sigma$ in relative entropy~(\ref{eq:red}) of fixed system dimension $m$ better reduces the fluctuation of relative entropy as a distinguishability measure, and make asymptotic discrimination performance more consistent across random state pairs, as reflected in the smaller variation of their optimal discrimination exponents. Secondly, the different finite variance expressions~(\ref{eq:r1}), (\ref{eq:r2}), (\ref{eq:r3}) and their numerical simulations further distinct the HS ensemble, BH ensemble, and inter-ensemble from each other, namely, by comparing at the same $m$ and fixed parameters, the HS ensemble with relatively smallest variance~(\ref{eq:r1}) and the BH ensemble has relatively largest variance~(\ref{eq:r2}), where as the inter-ensemble~(\ref{eq:r3}) lies in between while behaves closely to the HS ensemble, indicating a possibly similar behavior to the HS ensemble for the inter-ensemble that approximates the full probabilistic model BH ensemble by a less informative reduced-order approximate model HS ensemble. On the other hand, the Corollary~\ref{vanish} establishes the typicality of relative entropy under asymptotic regime~(\ref{eq:limfixn-m}) for all three cases of HS ensemble, BH ensemble, and inter-ensemble, under the fixed dimension difference assumption $F$, suggesting that relative entropy as a statistical quantity is concentrated to the mean for the generic random states studied, making the ensemble averages predictive for individual realizations.

\section{Derivation of variance formulas}\label{sec:calc}
In this section, we present the derivation of the variance results in the previous section. We first introduce the formulation of the two random state ensembles in Section~\ref{appd:ensembles}. We then proceed with the calculation of cumulant integrals over such ensembles in Section~\ref{sec:zon}. The strategy of calculating these integrals is that we first perform a change of variables, and apply a unique decomposition of positive integer power of trace of Hermitian matrix into zonal polynomials, which breaks calculation of joint cumulants into single ones. We then give the proof of the each propositions in Section~\ref{sec:prhs}, \ref{sec:prbh}, \ref{sec:prhsbh} by performing the calculation of the required terms over the corresponding ensemble using eigenvalue correlation functions after the decompositions. We also perform the proof of the asymptotic results by applying the limiting behavior of polygamma and finite summation functions in Section~\ref{sec:prcoro}.
\subsection{Random states ensembles}\label{appd:ensembles}
The two representative random states ensemble considered in this work is the HS~(\ref{eq:HS}) and BH~(\ref{eq:BH}) ensembles. The HS ensemble represents the simplest model for generic quantum states, requiring no prior assumptions about the states themselves. Its randomness arises from coefficients that are assumed to be independent and identically Gaussian distributed, corresponding to the least informative and most neutral probability distribution. As such, the HS ensemble can be viewed as the quantum analogue of the baseline Gaussian model that is widely used in statistical modeling of unknown variables. In studies of quantum information processing, generic random states drawn from the HS ensemble are commonly employed as benchmark states for evaluating performance~\cite{Page93,VPO16,Wei17,Wei20}. On the other hand, the BH ensemble is a refined version of the HS ensemble that satisfies several additional desirable properties. It is induced by the Bures metric, which is uniquely characterized as the only monotone metric that is both Fisher adjusted and Fubini–Study adjusted~\cite{Bures1969AnEO,Hall98,Zyczkowski01,Sommers03,Sommers04,Osipov10}. The corresponding distribution, often referred to as the Bures prior, is widely used in quantum state reconstruction from measurement data. Like states from the HS ensemble, generic states sampled from the BH ensemble are physically meaningful and can be generated efficiently in polynomial time.

We start the definition of the two ensembles by the formalism of bipartite system. A generic state of a bipartite system consists of two subsystems $A$ and $B$ of Hilbert space with dimensions $m$ and $n$, respectively, is defined as~\cite{BZ06}
\begin{equation}\label{eq:state}
\Ket{\psi}=\sum_{i=1}^{m}\sum_{j=1}^{n}c_{i,j}\Ket{i_{\rm{A}}}\otimes\Ket{j_{\rm{B}}},
\end{equation}
where coefficients $c_{i,j}$ are independent and identically distributed standard complex Gaussian random variables. Without loss of generality, we assume $m\leq n$, i.e. $A$ is the smaller subsystem. In the generic state ensemble obtained from the bipartite model~(\ref{eq:state}), $m$ denotes the system dimension, while $n$ is the environment (ancilla) dimension.

The HS ensemble has probability density function of the $m\times m$ reduced density matrix $\rho_{\rm{A}}=\tr_{\rm{B}}(\rho)$, given by partial tracing density matrix $\rho=\ket{\psi}\bra{\psi}$~\cite{Page93,BZ06},
\begin{equation}\label{eq:HS}
f_{\rm{HS}}\!\left(\rho_{\rm{A}}\right)=\frac{1}{C_{\rm{HS}}}
\delta\left(1-\tr\rho_{\rm{A}}\right)\det\!\!\!~^{\alpha}(\rho_{\rm{A}})\mathbbm{1}_{\{\rho_{\mathrm A}\succ 0\}},
\end{equation}
where $\delta(x)$ is the Dirac delta function, with the constant
\begin{equation}\label{eq:cHS}
C_{\rm{HS}}=\frac{\pi^{\frac{1}{2}m(m-1)}}{\Gamma\left(m(m+\alpha)\right)}\prod_{i=1}^{m}\Gamma(i+\alpha),
\end{equation}
where $\Gamma(x)$ is the Gamma function, and
\begin{equation}\label{eq:a}
\alpha=n-m\geq0.
\end{equation}
The matrix model of HS ensemble~(\ref{eq:HS}) is the fixed-trace Wishart ensemble~\cite{Page93,Ruiz95,huang2025cumulantstructuresentanglemententropy} constructed from a Wishart matrix $YY^{\dag}$ as
\begin{equation}\label{eq:maHS}
\rho_{\rm{A}}=\frac{YY^{\dag}}{\tr\left(YY^{\dag}\right)}
\end{equation}
with $Y$ denoting an $m\times n$ matrix of independent complex Gaussian entries.

The BH ensemble is a generalization of the HS ensemble~(\ref{eq:HS}) constructed by rotating the the generic state (\ref{eq:state}) by an $m\times m$ random unitary matrix $U$,
\begin{equation}\label{eq:BHs}
\Ket{\varphi}\propto\Ket{\psi}+\left(U\otimes I_{n}\right)\Ket{\psi}
\end{equation}
where $I_{n}$ denotes an identity matrix of dimension $n$. The probability density function of the unitary matrix $U$ is proportional to $\left|\det\left(I_{m}+U\right)\right|^{2\alpha}$. The Bures-Hall ensemble has probability density function of $\rho_{\rm{A}}=\tr_{\rm{B}}(\rho)$, given by partial tracing density matrix $\rho=\ket{\varphi}\bra{\varphi}$~\cite{Sarkar19,BZ06},
\begin{equation}\label{eq:BH}
f_{\rm{BH}}\!\left(\rho_{\rm{A}}\right)=\frac{1}{C_{\rm{BH}}}
\delta\left(1-\tr\rho_{\rm{A}}\right)\det\!\!\!~^{\alpha}(\rho_{\rm{A}})\!\!\int_{Z}\!\!\e^{-\!\tr\left(\rho_{\rm{A}}Z^{2}\right)} \dd Z\mathbbm{1}_{\{\rho_{\mathrm A}\succ 0\}},
\end{equation}
where
\begin{equation}\label{eq:cBH}
C_{\rm{BH}}=\frac{\pi^{m^2}m!~\!2^{-m(m+2\alpha-1)}}{\Gamma\left(m(m+2\alpha)/2\right)}\prod_{i=1}^{m}\frac{\Gamma(i+2\alpha)}{\Gamma(i+\alpha)},
\end{equation}
and $Z$ is an $m\times m$ Hermitian matrix. The matrix model corresponding to the state~(\ref{eq:BHs}) of Bures-Hall ensemble is~\cite{Sarkar19}
\begin{equation}\label{eq:maBH}
\rho_{\rm{A}}=\frac{\left(I_{m}+U\right)YY^{\dag}\left(I_{m}+U^{\dag}\right)}{\tr\left(\left(I_{m}+U\right)YY^{\dag}\left(I_{m}+U^{\dag}\right)\right)},
\end{equation}
where $YY^{\dag}$ is a Wishart matrix as in~(\ref{eq:maHS}).

The Dirac delta function $\delta$ in the ensembles~(\ref{eq:HS}) and~(\ref{eq:BH}) is a reflection of the defining property of density matrices~(\ref{eq:dm}) that
\begin{equation}\label{eq:t1}
\tr\rho_{\rm{A}}=1.
\end{equation}

\subsection{Calculation of moment integrals via zonal polynomials}\label{sec:zon}
The derivation of the variance formulas of relative entropy~(\ref{eq:r1}),~(\ref{eq:r2}),~(\ref{eq:r3}) requires the calculation of $\mathbb{E}[\operatorname{tr}(\rho \ln \sigma)]$, $\mathbb{E}[\operatorname{tr}(\rho \ln \rho) \operatorname{tr}(\rho \ln \sigma)]$ and $\mathbb{V}[\operatorname{tr}(\rho \ln \sigma)]$ in~(\ref{eq:rev4}), which needs to be decomposed into the moments/cumulants of linear statistics of single density matrix $\rho$ or $\sigma$, respectively. In the literature~\cite{wei2025averagerelativeentropyrandom}, a result
\begin{eqnarray}\label{eq:rhomsig}
    \mathbb{E}[\operatorname{tr}(\rho \ln \sigma)]=\frac{1}{m} \mathbb{E}[\operatorname{tr}( \ln \sigma)]
\end{eqnarray}
has been proved and obtained Loc. cit. for HS~(\ref{eq:HS}) and BH ensemble~(\ref{eq:BH}). The key to the derivation of~(\ref{eq:rhomsig}) is the connection of cumulant integrals to zonal polynomials~\cite{Forrester,Khatri66,Mathai}, and we apply such method further to the computation of desired higher-order moments  $\mathbb{E}[\operatorname{tr}(\rho \ln \rho) \operatorname{tr}(\rho \ln \sigma)]$ and variance $\mathbb{V}[\operatorname{tr}(\rho \ln \sigma)]$ in~(\ref{eq:rev4}), where the latter one is performed over $\mathbb{E}\!\left[\operatorname{tr}^2(\rho \ln \sigma)\right]$ via the relation
\begin{eqnarray}\label{eq:Ve2rhosig}
    \mathbb{V}[\operatorname{tr}(\rho \ln \sigma)]&=&\mathbb{E}\!\left[\operatorname{tr}^2(\rho \ln \sigma)\right]-\mathbb{E}^2\!\left[\operatorname{tr}(\rho \ln \sigma)\right]\\
    &=&\mathbb{E}\!\left[\operatorname{tr}^2(\rho \ln \sigma)\right]-\frac{1}{m^2} \mathbb{E}^2[\operatorname{tr}( \ln \sigma)].
\end{eqnarray}
We first introduce a change of variables. Let
    \begin{eqnarray} \label{eq:cov}
    X=r_1\rho,~~~~Y=r_2\sigma,
\end{eqnarray}
 where
    \begin{eqnarray}
r_1= \operatorname{tr}(X), ~~~~r_2=\operatorname{tr} (Y).
\end{eqnarray}
 The probability density function of $r$ is clearly
\begin{equation}\label{eq:hr}
h(r_i)=\frac{1}{\Gamma(d_i)} e^{-r_i} r_i^{d_i-1}, \quad i=1,2,\end{equation}
where\begin{eqnarray}\label{eq:di}
    d_i= \begin{cases}m n_i & \text{Hilbert-Schmidt} \\ \frac{m}{2}(2 n_i-m) & \text{Bures-Hall},\end{cases} \quad i=1,2.
\end{eqnarray}
It is also straightforward from density functions of $r_i$ that
\begin{eqnarray}
% \mathbb{E}\!\left[r_1\right]&=&d_1 \label{eq:r1}\\
\mathbb{E}\!\left[r_1^2\right]&=&d_1(d_1+1) \label{eq:r1^2}\\
\mathbb{E}\!\left[r_1^2 \ln r_1 \right]&=&d_1(d_1+1)\psi_0(d_1+2) \label{eq:r1^2lnr1}\\
\mathbb{E}\!\left[\ln r_2\right]&=&\psi_0(d_2) \label{eq:lnr2}\\
\mathbb{E}\!\left[\ln^2 r_2\right]&=&\psi_0^2(d_2)+\psi_1(d_2).\label{eq:ln^2r2}
\end{eqnarray}

Apply change of variables~(\ref{eq:cov}) to $\mathbb{E}[\operatorname{tr}(\rho \ln \rho) \operatorname{tr}(\rho \ln \sigma)]$ and  $\mathbb{E}[\operatorname{tr}^2(\rho \ln \sigma)]$, we deduce that
\begin{eqnarray}\label{eq:trrholnrhotrrholnsigma}
&&\mathbb{E}[\operatorname{tr}(\rho \ln \rho) \operatorname{tr}(\rho \ln \sigma)]\nonumber\\
&=&\frac{1}{\mathbb{E}\left[r_1^2\right]} \mathbb{E}\left[\operatorname{tr}(X \ln X)\operatorname{tr}(X \ln Y)\right]-\mathbb{E}[\ln r_2]\mathbb{E}[\operatorname{tr}(\rho \ln \rho)]
-\frac{\mathbb{E}\left[r_1^2 \ln r_1\right]}{\mathbb{E}\left[r_1^2\right]} \left(\mathbb{E}[\operatorname{tr}\left(\rho \ln \sigma\right)]+\mathbb{E}[\ln r_2]\right),
\end{eqnarray}
and
\begin{eqnarray}\label{eq:tr^2rhosigma}
\mathbb{E}\!\left[\operatorname{tr}^2(\rho \ln \sigma)\right]
&=&\frac{1}{\mathbb{E}\left[r_1^2\right]} \mathbb{E}\!\left[\operatorname{tr}^2(X \ln Y)\right]\nonumber-2 \mathbb{E}[\ln r_2] \mathbb{E}[\operatorname{tr}(\rho \ln \sigma)]-\mathbb{E}[\ln^2 r_2],
\end{eqnarray}
respectively. Hence the calculation now boils down to $\mathbb{E}\left[\operatorname{tr}(X \ln X)\operatorname{tr}(X \ln Y)\right]$ and $\mathbb{E}\!\left[\operatorname{tr}^2(X \ln Y\right]$, which requires a machinery to decompose integrals over $\operatorname{tr}^l(X \ln Y)$ into zonal polynomials introduced as follows.

For an arbitrary Hermitian matrix $X$, we have a unique decomposition on the $l$-th positive integer power of the trace~\cite{wei2025averagerelativeentropyrandom,Mathai}
\begin{equation}\label{eq:trc}
\tr^{l}\!\left(X\right)=\sum_{\kappa}C_{\kappa}(X),
\end{equation}
where $C_{\kappa}(X)$ are zonal polynomials, with summation taken over partitions $\kappa=(\kappa_{1},\kappa_{2},\dots,\kappa_{m})$ of $l$ into no more than $m$ parts,
\begin{equation}
\kappa_{1}+\kappa_{2}+\dots+\kappa_{m}=l,~~~~~~\kappa_{1}\geq\kappa_{2}\geq\dots\geq\kappa_{m}\geq0.
\end{equation}
Let $\{x_i\}_{i=1}^{m}$ be the $m$ eigenvalues of $X$, the zonal polynomial $C_{\kappa}(X)$ is a symmetric polynomial given by the representation theory of symmetric group~\cite{James64},
\begin{equation}\label{eq:czp}
C_{\kappa}(X)=\chi_{\kappa}(1)\chi_{\kappa}(X),
\end{equation}
where
\begin{equation}\label{eq:czp1}
\chi_{\kappa}(1)=\frac{l!\prod_{1\leq i<j\leq m}(\kappa_{i}-\kappa_{j}-i+j)}{\prod_{j=1}^{m}(\kappa_{j}+m-j)!}
\end{equation}
is the dimension of representation of the symmetric group, and
\begin{equation}\label{eq:czp2}
\chi_{\kappa}(X)=\frac{\det\left(x_{i}^{\kappa_{j}+m-j}\right)}{\det\left(x_{i}^{m-j}\right)}
\end{equation}
is character of the representation, which is a symmetric function in $\{x_i\}_{i=1}^{m}$.
The character~(\ref{eq:czp2}) is also known as the Schur polynomial, which admits a representation by determinant of elementary symmetric polynomials~\cite{Macdonald},
\begin{equation}\label{eq:czp3}
\chi_{\kappa}(X)=\det\left(e_{\kappa'-i+j}\left(x_{1},x_{2},\dots,x_{m}\right)\right)_{i,j=1}^{l(\kappa')},
\end{equation}
where $l(\kappa)$ and $\kappa'$ denote the length and the conjugate of a partition $\kappa$, respectively, and
\begin{equation}
e_{k}\left(x_{1},\dots,x_{m}\right)=\sum_{1\leq i_1<\cdots<i_k\leq m}x_{i_1}x_{i_2}\cdots x_{i_k}.
\end{equation}

For the matrices $X$ and $Y$ defined in (\ref{eq:cov}), which are clearly Hermitian where $\rho$ and $\sigma$ are Hermitian random ensembles, we consider eigenvalue decompositions of the density matrices
\begin{equation}
X=V\Sigma_{x}V^{\dag},~~~~~~~~Y=W\Sigma_{y}W^{\dag},
\end{equation}
where the diagonal matrices $\Sigma_{x}$, $\Sigma_{y}$ consist of eigenvalues of $X$ and $Y$, respectively, and $V$, $W$ are unitary matrices of the eigenvectors, respectively. The densities~(\ref{eq:HS}) and~(\ref{eq:BH}) are unitarily invariant, hence so are the density functions of $X$ and $Y$ from the HS and BH ensembles considered in this work. Similar to~\cite{wei2025averagerelativeentropyrandom}, the required moment integral terms on
\begin{eqnarray}\label{eq:rhosigdepend}
    \operatorname{tr}^l(X\ln Y)=\operatorname{tr}^l(\Sigma_x U \ln \Sigma_y U^{\dag})
\end{eqnarray}
do depend on the eigenvectors through the unitary matrix $U=V^{\dag}W$, while the other terms on $\operatorname{tr}^l(X\ln X)$ do not depend on the eigenvectors as
\begin{eqnarray}
    \operatorname{tr}^l(X\ln X) = \operatorname{tr}^l(\Sigma_{x} \ln \Sigma_{x}).
\end{eqnarray}
Since the densities~(\ref{eq:HS}) and~(\ref{eq:BH}) are unitarily invariant, so are the Hermitian matrices $X$ and $Y$. Therefore, the moment integrals of~(\ref{eq:rhosigdepend}) over eigenvalues
and eigenvectors can be calculated separately, which is performed via the key result on the integral identity of zonal polynomials as follows.\\

For two $m\times m$ Hermitian matrices $X$, $Y$ over the invariant measure $\dd U$ of the unitary group $U(m)$, an integral identity of zonal polynomials~\cite{wei2025averagerelativeentropyrandom,Khatri66,Forrester}
\begin{equation}\label{eq:AUBU}
\int_{U(m)}C_{\kappa}\!\left(XUYU^{\dag}\right)\dd U=\frac{C_{\kappa}\!\left(X\right)C_{\kappa}\!\left(Y\right)}{C_{\kappa}\!\left(I_{m}\right)}
\end{equation}
holds for any partition $\kappa$. As a result, for $l=1$ in~(\ref{eq:trc}),
\begin{eqnarray}\label{decom:trxlny}
    \int_{U(m)}\tr(X\ln Y)\dd U &=&  \int_{U(m)}C_{(1)}(\Sigma_x U \ln \Sigma_y U^{\dag}) \dd U\nonumber\\
    &=& \frac{C_{(1)}(\Sigma_x)C_{(1)}(\ln \Sigma_y)}{C_{(1)}(I_m)}\nonumber \\
     &=&\frac{1}{m}\operatorname{tr}(X)\operatorname{tr} (\ln Y).
\end{eqnarray}
where the only partition on the right-hand side of~(\ref{eq:trc}) is $\kappa=(1)$ denoting
\begin{equation}\label{eq:k1}
\kappa_{1}=1,~~~~\kappa_{2}=\dots=\kappa_{m}=0.
\end{equation}
From~(\ref{eq:czp3}) we also calculate that for identity matrix $I_m$,
\begin{eqnarray}
\chi_{(1)}(I_m)&=&e_1(1,\cdots,1) =  m,
\end{eqnarray}
which yields
\begin{eqnarray}
     C_{(1)}(I_m)=\chi_{(1)}(1)\chi_{(1)}(I_m)=m \label{eq:C1Im}
 \end{eqnarray}
 via~(\ref{eq:czp}).
Similarly, for $l=2$ in~(\ref{eq:trc}), the two possible partitions on the right-hand side of~(\ref{eq:trc}) is $\kappa=(2,0)$ denoting
\begin{equation}\label{eq:k20}
\kappa_{1}=2,~~~~\kappa_{2}=\dots=\kappa_{m}=0,
\end{equation}
and
$\kappa=(1,1)$ denoting
\begin{equation}\label{eq:k11}
\kappa_{1}=1,~~~~ \kappa_{2}=1, ~~~~\kappa_{3}=\dots=\kappa_{m}=0.
\end{equation} Therefore, applying the result~(\ref{eq:AUBU}) to the second-moment term yields
\begin{eqnarray}
    \int_{U(m)}\tr^2(X\ln Y)\dd U &=&  \int_{U(m)}C_{(2,0)}(\Sigma_x U \ln \Sigma_y U^{\dag}) \dd U+ \int_{U(m)}C_{(1,1)}(\Sigma_x U \ln \Sigma_y U^{\dag}) \dd U\nonumber\\
    &=& \frac{C_{(2,0)}(\Sigma_x)C_{(2,0)}(\ln \Sigma_y)}{C_{(2,0)}(I_m)}+\frac{C_{(1,1)}(\Sigma_x)C_{(1,1)}(\ln \Sigma_y)}{C_{(1,1)}(I_m)}\nonumber\\
    &=& \frac{2 C_{(2,0)}(X)C_{(2,0)}(\ln Y)}{m(m+1)}+\frac{2 C_{(1,1)}(X)C_{(1,1)}(\ln Y)}{m(m-1)} ,\label{decom:tr^2xlny}
\end{eqnarray}
where we have used that from~(\ref{eq:czp3}),
\begin{eqnarray}
    \chi_{(2,0)}(I_m)&=&e_1^2(1,\cdots,1)-e_2(1,\cdots,1) = \frac{1}{2}m(m+1)\\
   \chi_{(1,1)}(I_m)&=&e_2(1,\cdots,1) = \frac{1}{2}m(m-1),
\end{eqnarray}
and consequently the zonal polynomials for the above cases
\begin{eqnarray}
    C_{(2,0)}(I_m)&=&\chi_{(2,0)}(1) \chi_{(2,0)}(I_m)=\frac{1}{2}m(m+1) \label{eq:C2Im}\\
    C_{(1,1)}(I_m)&=&\chi_{(1,1)}(1) \chi_{(1,1)}(I_m)=\frac{1}{2}m(m-1) \label{eq:C1^2Im}.
\end{eqnarray}
We have also adopted the fact that
\begin{eqnarray}
    C_{\kappa}(\Sigma_x)= C_{\kappa}(V\Sigma_x V^{\dag})= C_{\kappa}(X)
\end{eqnarray}
for arbitrary Hermitian matrix $X$ and partition $\kappa$ since zonal polynomials are unitary invariant~\cite{James64}.\\

As a result of~(\ref{decom:trxlny}), we have
\begin{eqnarray}
    \mathbb{E}\!\left[\operatorname{tr}(X\ln Y)\right]=\frac{1}{m}\mathbb{E}\!\left[\operatorname{tr}(X)\right]\mathbb{E}\!\left[\operatorname{tr}(\ln Y)\right],
\end{eqnarray}
and similarly, the first desired term of moment integrals $\mathbb{E}\!\left[\operatorname{tr}(X\ln X)\operatorname{tr}(X\ln Y)\right]$ is decoupled as
\begin{eqnarray}
    \mathbb{E}\!\left[\operatorname{tr}(X\ln X)\operatorname{tr}(X\ln Y)\right]
    &=&\frac{1}{m}\mathbb{E}\!\left[\operatorname{tr}(X\ln X)\operatorname{tr}(X)\operatorname{tr}(\ln Y)\right]\nonumber\\
    &=&\frac{1}{m}\mathbb{E}\!\left[\operatorname{tr}(X\ln X)\operatorname{tr}(X)\right]\mathbb{E}\!\left[\operatorname{tr}(\ln Y)\right].\label{decom:xlnxxlny}
\end{eqnarray}
Moreover, for the first term in (\ref{decom:xlnxxlny}), we also have a simple decomposition result from~\cite{huang2025cumulantstructuresentanglemententropy} that
\begin{eqnarray}\label{decom:xlnxx}
\!\!\!\!\!\!\!\! \!\!\!\!\!\!\!\!  &&\!\!\!\! \mathbb{E}\!\left[\operatorname{tr}(X\ln X)\operatorname{tr}(X)\right]\nonumber\\
\!\!\!\! \!\!\!\! \!\!\!\! \!\!\!\! &=&\!\!\!\! \mathbb{E}\!\left[\operatorname{tr}(X\ln X)\right]+\mathbb{E}\!\left[\operatorname{tr}(X)\right]+\mathbb{E}\!\left[\operatorname{tr}(X\ln X)\right]\mathbb{E}\!\left[\operatorname{tr}(X)\right].\label{indv}
\end{eqnarray}
In the literature, for Hermitian matrix $X$ with parameters $m$ and $n$, the individual first-order moments $\mathbb{E}\!\left[\operatorname{tr}(X)\right]$ and $\mathbb{E}\!\left[\operatorname{tr}(X\ln X)\right]$ in~(\ref{indv}) have been obtained in~\cite{Page93,Foong94,Ruiz95,VPO16,huang2025cumulantstructuresentanglemententropy} and~\cite{Sarkar19,Wei20BHA,wei2026spectralmomentsbureshallensemble} for HS~(\ref{eq:HS}) and BH~(\ref{eq:BH}) ensemble, respectively,
\begin{eqnarray}
     \mathbb{E}\!\left[\operatorname{tr}(X_{\mathrm{HS}})\right]&=& m n \\
 \mathbb{E}\!\left[\operatorname{tr}(X_{\mathrm{BH}})\right]&=& \frac{1}{2} m (2n-m)\\
  \mathbb{E}\!\left[\operatorname{tr}(X_{\mathrm{HS}}\ln X_{\mathrm{HS}})\right]&=& m n\psi_0(n+1)+\frac{1}{2} m(m-1),\\
 \mathbb{E}\!\left[\operatorname{tr}(X_{\mathrm{BH}}\ln X_{\mathrm{BH}})\right]&=& \frac{1}{2} m (2n-m) \psi_0 \!\left(n+\frac{1}{2}\right),
\end{eqnarray}
while first-order moment corrsponding to the other term $\mathbb{E}\!\left[\operatorname{tr}(\ln Y)\right]$ in~(\ref{decom:xlnxxlny}) has been obtained in~\cite{huang2025cumulantstructuresentanglemententropy} and~\cite{wei2026spectralmomentsbureshallensemble} for HS and BH ensemble, respectively,
\begin{eqnarray}
     \mathbb{E}\!\left[\operatorname{tr}(\ln X_{\mathrm{HS}})\right]&=& n\psi_0 (n)-(n-m)\psi_0 (n-m)-m\\
 \mathbb{E}\!\left[\operatorname{tr}(\ln X_{\mathrm{BH}})\right]&=& 2(2 n-m) \psi_0(2 n-m)-(n-m)\left(4 \psi_0(2 n-2 m)-\psi_0(n-m)\right)\nonumber\\
 &&-n \psi_0(n) -m\left(\psi_0(2)-\psi_0\left(\frac{1}{2}\right)\right).
 \end{eqnarray}
Insert the required terms accordingly with $n$ evaluated as $n_1$ and $n_2$ for $X$ and $Y$ in~(\ref{eq:cov}), respectively, one completes the calculation of the desired term $\mathbb{E}\!\left[\operatorname{tr}(X\ln X)\operatorname{tr}(X\ln Y)\right]$ for each of the cases involving $X_{\mathrm{HS}}$, $X_{\mathrm{BH}}$, $Y_{\mathrm{HS}}$, and $Y_{\mathrm{BH}}$. Consequently, one obtains the result of $\mathbb{E}[\operatorname{tr}(\rho \ln \rho) \operatorname{tr}(\rho \ln \sigma)]$ via~(\ref{eq:trrholnrhotrrholnsigma}) for each of the cases involving $\rho_{\mathrm{HS}}$, $\rho_{\mathrm{BH}}$, $\sigma_{\mathrm{HS}}$, and $\sigma_{\mathrm{BH}}$ via~(\ref{eq:trrholnrhotrrholnsigma}). This addresses the first required term in the derivation of variance of relative entropy.

The bulk of the detailed calculation of specific moment integrals of this work then falls into the required term $ \mathbb{E}\!\left[\operatorname{tr}^2(X\ln Y)\right] $, where we have the decomposition via zonal polynomials from~(\ref{decom:tr^2xlny}),
\begin{eqnarray}
   \mathbb{E}\!\left[\operatorname{tr}^2(X\ln Y)\right] &=&\frac{2 \mathbb{E}\!\left[ C_{(2,0)}(X)\right] \mathbb{E}\!\left[C_{(2,0)}(\ln Y)\right]}{m(m+1)}+\frac{2  \mathbb{E}\!\left[C_{(1,1)}(X)\right]  \mathbb{E}\!\left[C_{(1,1)}(\ln Y)\right]}{m(m-1)},
   \label{tr^2xlny}
\end{eqnarray}
which is the starting point of the proof of the variance propositions.
Moreover, from~\cite{Khatri70} we have that for an arbitrary Hermitian matrix $X$,
\begin{eqnarray}
    \mathbb{E}\!\left[C_{(2,0)}(X)\right]=\mathbb{E}\!\left[\frac{R_1^2+R_2}{2}\right] &=&\frac{1}{2}(\mathbb{E}\!\left[R_1^2\right]+\mathbb{E}\!\left[R_2\right])\label{inttable1}\\
     \mathbb{E}\!\left[C_{(2,0)}(\ln X)\right]=\mathbb{E}\!\left[\frac{H_1^2+H_2}{2}\right]&=&\frac{1}{2}(\mathbb{E}\!\left[H_1^2\right]+\mathbb{E}\!\left[H_2\right])\label{inttable2}\\
      \mathbb{E}\!\left[C_{(1,1)}(X)\right]=\mathbb{E}\!\left[\frac{R_1^2-R_2}{2}\right]&=&\frac{1}{2}(\mathbb{E}\!\left[R_1^2\right]-\mathbb{E}\!\left[R_2\right])\label{inttable3}\\
      \mathbb{E}\!\left[C_{(1,1)}(\ln X)\right]=\mathbb{E}\!\left[\frac{H_1^2-H_2}{2}\right]&=&\frac{1}{2}(\mathbb{E}\!\left[H_1^2\right]-\mathbb{E}\!\left[H_2\right])\label{inttable4},
\end{eqnarray}
where
\begin{eqnarray}
    R_k=\sum_{i=1}^m x_i^k,~~~~H_k=\sum_{i=1}^m \ln^k x_i,
\end{eqnarray}
with $\{x_i\}_{i=1}^{m}$ denoting the eigenvalues of $X$. Among the right-hand sides of~(\ref{inttable1})--(\ref{inttable4}), $\mathbb{E}\!\left[R_2\right]$ have been obtained in~\cite{huang2025cumulantstructuresentanglemententropy} and~\cite{Sarkar19,LW21,wei2026spectralmomentsbureshallensemble} for HS~(\ref{eq:HS}) and BH~(\ref{eq:BH}) ensemble, respectively. As for $\mathbb{E}\!\left[R_1^2\right]$, one notices that for $X$ in~(\ref{eq:cov}),
\begin{eqnarray}
    R_1=\operatorname{tr}(X)=r_1
\end{eqnarray}
in~(\ref{eq:hr}), hence
\begin{eqnarray}
    \mathbb{E}\!\left[R_1^2\right]&=&d_1(d_1+1)
\end{eqnarray}
for $d_1$ in~(\ref{eq:di}), and similarly for the $R_k$ over $Y$ in~(\ref{eq:cov}).
The rest of the calculation is to compute $\mathbb{E}\!\left[H_1^2\right]$, $\mathbb{E}\!\left[H^2\right]$ for HS~(\ref{eq:HS}) and BH~(\ref{eq:BH}) ensemble, according to the corresponding relative entropy.

\subsection{Proof of Proposition~\ref{prop:HS}}\label{sec:prhs}
\begin{proof}
The remaining task for variance of relative entropy over HS ensemble~(\ref{eq:HS}) is to calculate the moments of $H_k$, which is known to be linear statistics over Laguerre ensemble~\cite{Page93,Wei17,huang2025cumulantstructuresentanglemententropy}, whose correlation kernel is
\begin{eqnarray}
    K(x, y)=\sqrt{w(x) w(y)} \sum_{k=0}^{m-1} \frac{k!}{(k+\alpha)!} L_k^{(\alpha)}(x) L_k^{(\alpha)}(y) ,
\end{eqnarray}
where
\begin{eqnarray}\label{eq:lg}
    L_k^{(\alpha)}(x)=\sum_{i=0}^k(-1)^i\binom{\alpha+k}{k-i} \frac{x^i}{i!}
\end{eqnarray} is the (generalized) Laguerre orthogonal polynomial with weight
\begin{eqnarray}
    w(x)=\e^{-x}x^{\alpha}.
\end{eqnarray}
The calculation of $\mathbb{E}\!\left[H_2\right]$ follows by taking derivatives in the calculation of
\begin{eqnarray}
    \mathbb{E}\!\left[R_t\right] =m \int_0^{\infty} x^t  g_1\!\left(x\right) \mathrm{d} x,
\end{eqnarray}
performed over the one-point correlation kernel
 \begin{eqnarray}
   g_1\!\left(x\right) & =&\frac{1}{m} K\!\left(x, x\right),
 \end{eqnarray}
and utilize the explicit summation representation~(\ref{eq:lg}), which results in
\begin{eqnarray}
\mathbb{E}\!\left[H_2\right]&=&\left.\frac{\mathrm{d}^2}{\mathrm{d} t^2}\mathbb{E}\!\left[R_t\right] \right|_{t=0}\nonumber\\&=&2m \Psi_{0,\alpha}+\frac{1}{n}(-2 \left(m n+m-n^2\right) \psi_0 (n-m)-2  \left(n (m-n) \psi_0 (n-m)-m+n^2\right)\psi_0 (n)\nonumber\\&&+2 n (m-n) \psi_0^2 (n-m)+m n \psi_0^2 (n)+m n \left(\psi _1(n)+2\right).
\end{eqnarray}
We also evaluate $\alpha$ as $\alpha_1$, $\alpha_2$, and $n$ as $n_1$, $n_2$ for $X$, $Y$ in~(\ref{eq:cov}) as needed in~(\ref{tr^2xlny}).
 % The calculation of $\mathbb{E}\!\left[R_1^2\right]$ utilizes correlation kernels of Laguerre ensemble, denoted $g_1(x)$ and $g_2(x,y)$,
% \begin{eqnarray}
%     \mathbb{E}\!\left[R_1^2\right]&=&
% m \int_0^{\infty} x^2  g_1\!\left(x\right) \mathrm{d} x+m(m-1) \int_0^{\infty}\int_0^{\infty} x y g_2\!\left(x, y\right) \mathrm{d} x \mathrm{d} y ,
% \end{eqnarray}
% where
% \begin{eqnarray}
%     \!\!\!\!g_1\!\left(x\right) &\!\!\!\! =\!\!\!\!&\frac{1}{m} K\!\left(x, x\right), \\
% \!\!\!\!g_2\!\left(x, y\right) &\!\!\!\! =\!\!\!\!&\frac{1}{m(m-1)}\left(K\!\left(x, x\right) K\!\left(y, y\right)-K^2\!\left(x, y\right)\right) .
% \end{eqnarray}
 On the other hand,
\begin{eqnarray}
\mathbb{E}\!\left[H_1^2\right]=\kappa_2\left(H_1\right)+\kappa_1^2\left(H_1\right),
\end{eqnarray}
where we have from~\cite{huang2025cumulantstructuresentanglemententropy} for the HS~(\ref{eq:HS}) case, \begin{eqnarray}
 \kappa_1\left(H_1\right)&=&n \psi_0(n)-(n-m) \psi_0(n-m)-m\\
    \kappa_2\left(H_1\right)&=&n \psi_1(n)-(n-m) \psi_1(n-m)+\psi_0(n)-\psi_0(n-m),
\end{eqnarray}
and we again evaluate $n$ as $n_1$, $n_2$ for $X$, $Y$ in~(\ref{eq:cov}) as needed in~(\ref{tr^2xlny}).

The above calculation lead to the results of $\mathbb{E}\!\left[C_{(2,0)}(X_{\mathrm{HS}})\right]$, $\mathbb{E}\!\left[C_{(2,0)}(\ln X_{\mathrm{HS}})\right]$, $\mathbb{E}\!\left[C_{(1,1)}(X_{\mathrm{HS}})\right]$, $\mathbb{E}\!\left[C_{(1,1)}(\ln X_{\mathrm{HS}})\right]$ via~(\ref{inttable1})--(\ref{inttable4}), and similarly for the moments over $Y_{\mathrm{HS}}$ in~(\ref{eq:cov}) by replacing $n_1$ by $n_2$ accordingly. This completes the calculation of the desired term $\mathbb{E}\!\left[\operatorname{tr}^2(X_{\mathrm{HS}}\ln Y_{\mathrm{HS}})\right]$ via~(\ref{decom:tr^2xlny}), and consequently the required $\mathbb{V}[\operatorname{tr}(\rho_{\mathrm{HS}} \ln \sigma_{\mathrm{HS}})]$ via~(\ref{eq:tr^2rhosigma}) and~(\ref{eq:Ve2rhosig}). Together with~(\ref{eq:rhomsig}), the derived result on $\mathbb{E}[\operatorname{tr}(\rho_{\mathrm{HS}} \ln \rho_{\mathrm{HS}}) \operatorname{tr}(\rho_{\mathrm{HS}} \ln \sigma_{\mathrm{HS}})]$ via~(\ref{decom:xlnxxlny}), (\ref{decom:xlnxx}) and (\ref{eq:trrholnrhotrrholnsigma}), and known results on $\mathbb{E}\!\left[\operatorname{tr}(\rho_{\mathrm{HS}}\ln \rho_{\mathrm{HS}})\right]$ in~(\ref{EHS}) and $\mathbb{V}\!\left[\operatorname{tr}(\rho_{\mathrm{HS}}\ln \rho_{\mathrm{HS}})\right]$ in!(\ref{VHS}), respectively, we deduce the result in~(\ref{eq:r1}) via~(\ref{eq:rev4}). To obtain~(\ref{eq:degenr1}) from~(\ref{eq:r1}), we utilize the resolving the indeterminacy of polygamma functions with negative arguments in the limit $\epsilon\to0$ as~\cite{Wei20,Wei20BH},
\begin{subequations}\label{eq:pgna}
\begin{eqnarray}
\Gamma(-l+\epsilon)& =&\frac{(-1)^l}{l!\epsilon}\left(1+\psi_0(l+1) \epsilon+o\left(\epsilon^2\right)\right) \\
\psi_0(-l+\epsilon)&=&-\frac{1}{\epsilon}+\psi_0(l+1)+\left(2 \psi_1(1)-\psi_1(l+1)\right)\epsilon \nonumber\\
&&+\frac{1}{2}\psi_2(l+1) \epsilon^2+o\left(\epsilon^3\right) \\
\psi_1(-l+\epsilon)&=&\frac{1}{\epsilon^2}-\psi_1(l+1)+\psi_1(1)+\zeta(2)+o(\epsilon),
\end{eqnarray}
\end{subequations}
where $\zeta(s)$ is in~(\ref{zeta}).
\end{proof}

\subsection{Proof of Proposition~\ref{prop:BH}}\label{sec:prbh}
\begin{proof}
Similar calculation is performed for BH ensemble~(\ref{eq:BH}) with $R_k$ and $H_k$ over Cauchy-Laguerre biorthogonal ensembles~\cite{Wei20BHA,Wei20BH,LW21,wei2025skewnessvonneumannentropy,wei2026spectralmomentsbureshallensemble}, where the correlation kernels~\cite{FK16, LW21} are denoted as $K_{00}(x,y)$, $K_{01}(x,y)$, $K_{10}(x,y)$, and $K_{11}(x,y)$, which admit finite summation representation~\cite{FK16,Bertola14} as below, where
\begin{eqnarray}\label{eq:betanm}
    \beta=n-m-\frac{1}{2}=\alpha-\frac{1}{2}
\end{eqnarray} as in~(\ref{eq:beta}),
\begin{subequations}\label{kernels}
\begin{eqnarray}
    K_{00}(x,y)&=&\sum _{k=0}^{m-1} \frac{1}{h_k}p_k(x) q_k(y) \label{eq:00}\\
    K_{01}(x,y)&=&-x^{\beta }\e^{-x}\sum _{k=0}^{m-1}\frac{1}{h_k} p_k(y) Q_k(-x) \label{eq:01}\\
   K_{10}(x,y)&=&-y^{\beta +1}\e^{-y}\sum _{k=0}^{m-1} \frac{1}{h_k}P_k(-y) q_k(x) \label{eq:10}\\
    K_{11}(x,y)&=&x^{\beta }y^{\beta +1} \e^{-x-y}  \sum _{k=0}^{m-1}\frac{1}{h_k} P_k(-y) Q_k(-x)-W(x,y), \label{eq:11}
    \end{eqnarray}
    \end{subequations}
  where $p_{k}(x)$ and $q_{k}(y)$ are Cauchy-Laguerre biorthogonal polynomials,
\begin{equation}\label{eq:oc}
\int_{0}^{\infty}\!\!\int_{0}^{\infty}p_{k}(x)q_{l}(y)W(x,y)\dd x\dd y=h_k\delta_{k,l}
\end{equation}
with the normalizing constant $h_k$ and weight function
\begin{equation}
    W(x,y)= \frac{x^{\beta } y^{\beta +1} \e^{-x-y}}{x+y}.
\end{equation}
Also, the Cauchy transforms of $p_{k}(x)$ and $q_{k}(y)$ are respectively
\begin{subequations}
\begin{eqnarray}\label{eq:PQ}
P_{k}(x)&=&\int_{0}^{\infty}\frac{v^{\beta}\e^{-v}}{x-v}p_{k}(v)\dd v \\
Q_{k}(y)&=&\int_{0}^{\infty}\frac{w^{\beta+1}\e^{-w}}{y-w}q_{k}(w)\dd w.
\end{eqnarray}
\end{subequations}
We utilize their representations via Meijer G-functions~\cite{Bertola10,Bertola14,Forrester}
\begin{subequations}\label{eq:kerM}
\begin{eqnarray}
p_{k}(x)&=&\sqrt{2}(-1)^{k}G_{2,3}^{1,1}\left(\!\begin{array}{c}-2\beta-1-k;k+1\\0;-\beta,-2\beta-1\end{array}\Big|x\Big.\right)\\
q_{k}(x)&=&\sqrt{2}(-1)^{k}(\beta+k+1)G_{2,3}^{1,1}\left(\!\begin{array}{c}-2\beta-1-k;k+1\\0;-\beta-1,-2\beta-1\end{array}\Big|x\Big.\right)\\
P_{k}(x)&=&\sqrt{2}(-1)^{k+1}\e^{-x}G_{2,3}^{2,1}\left(\!\begin{array}{c}-\beta-k-1;\beta+k+1\\0,\beta;-\beta-1\end{array}\Big|-\!x\Big.\right)\\
Q_{k}(x)&=&\sqrt{2}(-1)^{k+1}(\beta+k+1)\e^{-x} G_{2,3}^{2,1}\left(\!\begin{array}{c}-\beta-k;\beta+k+2\\0,\beta+1;-\beta\end{array}\Big|-\!x\Big.\right),
\end{eqnarray}
\end{subequations}
which admit finite summation representations. In the same manner of~\cite{Wei20BH,LW21}, we consider the integrals of $\mathbb{E}\!\left[R_t\right]$, $\mathbb{E}\!\left[R_t^2\right]$ performed over the one-point and two-point kernels of the Cauchy-Laguerre ensemble, respectively,
\begin{eqnarray}
\mathbb{E}\!\left[R_t\right]&=& m\int_{0}^{\infty}\!\!x^{t} h_{1}(x)\mathrm{d} x\label{eq:ET1} \\
    \mathbb{E}\!\left[R_t^{2}\right]&=& m\int_{0}^{\infty}\!\!x^{2t} h_{1}(x)\mathrm{d} x+(m-1)_2\!\int_{0}^{\infty}\!\!\int_{0}^{\infty}\!\!x^ty^t h_{2}\!\left(x,y\right)\mathrm{d} x \mathrm{d} y, \\\label{eq:ET2}
\end{eqnarray}
where \begin{eqnarray}\label{eq:poch}
    (a)_n=\frac{\Gamma(a+n)}{\Gamma(a)}
\end{eqnarray} denotes the Pochhammer's symbol, and the density functions $h_1(x)$ and $ h_2(x,y)$ are~\cite{LW21,FK16}
\begin{eqnarray}
h_{1}(x)&=&\frac{1}{2m}\left(K_{01}(x,x)+K_{10}(x,x)\right)\label{eq:h1} \\
h_{2}(x,y)&=&\frac{1}{4m(m-1)}(\left(K_{01}(x,x)+K_{10}(x,x)\right)\left(K_{01}(y,y)+K_{10}(y,y)\right)\nonumber \\
&&-2K_{01}(x,y)K_{01}(y,x) -2K_{10}(x,y)K_{10}(y,x)\nonumber \\
&&-2K_{00}(x,y)K_{11}(x,y)-2K_{00}(y,x)K_{11}(y,x)).\label{eq:h2}
\end{eqnarray}
We illustrate the calculation of $\mathbb{E}\!\left[R_t^{2}\right]$ as follows, while $\mathbb{E}\!\left[R_t\right]$ is similar. In the same manner of~\cite{Wei20BH}, the integral of $\mathbb{E}\!\left[R_t^{2}\right]$ is divided into
\begin{equation}\label{eq:Tvi}
\mathbb{E}\!\left[R_t^{2}\right]=\mathbb{E}^2\!\left[R_t\right]+\frac{1}{2}\left(\mathrm{I_A}(t)-\mathrm{I_B}(t)-\mathrm{I_C}(t)-2\mathrm{I_D}(t)\right)
\end{equation}
via the symmetry, where
\begin{eqnarray}
\mathrm{I_A}(t)&=&\int_{0}^{\infty}x^{2t}\left(K_{01}(x,x)+K_{10}(x,x)\right)\dd x \label{eq:IA} \\
\mathrm{I_B}(t)&=&\int_{0}^{\infty}\!\!\int_{0}^{\infty}x^t y^t~K_{01}(x,y)K_{01}(y,x)\dd x\dd y \label{eq:IB} \\
\mathrm{I_C}(t)&=&\int_{0}^{\infty}\!\!\int_{0}^{\infty}x^t y^t~K_{10}(x,y)K_{10}(y,x)\dd x\dd y \label{eq:IC} \\
\mathrm{I_D}(t)&=&\int_{0}^{\infty}\!\!\int_{0}^{\infty}x^ty^t~K_{00}(x,y)K_{11}(x,y)\dd x\dd y\label{eq:ID}.
\end{eqnarray}
We apply the same method as in~\cite{Wei20BH} utilizing Mellin transforms to the individual integrals~(\ref{eq:IA}) and (\ref{eq:ID}), and integral identities for Meijer G-functions~\cite{Prudnikov86} (cf. (61) in~\cite{Wei20BH}) to~(\ref{eq:IB}), (\ref{eq:IC}), which result in finite summations as in~\cite{Wei20BH,wei2025skewnessvonneumannentropy}, and we list them in Appendix~\ref{appd:finsum}.
It is then convenient to calculate $\mathbb{E}\!\left[H_1^2\right]$ and $\mathbb{E}\!\left[H_2\right]$ as derivatives,
\begin{eqnarray}
\mathbb{E}\!\left[H_1^2\right]=\left.\frac{\mathrm{d}}{\mathrm{d} t}\mathbb{E}\!\left[R_t^2\right] \right|_{t=0},~~~~\mathbb{E}\!\left[H_2\right]=\left.\frac{\mathrm{d}^2}{\mathrm{d} t^2}\mathbb{E}\!\left[R_t\right] \right|_{t=0},
\end{eqnarray}
for which we omit presenting the expression of the intermediate result here for simplicity. The above results lead to $\mathbb{E}\!\left[C_{(2,0)}(X_{\mathrm{BH}})\right]$, $\mathbb{E}\!\left[C_{(2,0)}(\ln X_{\mathrm{BH}})\right]$,$\mathbb{E}\!\left[C_{(1,1)}(X_{\mathrm{BH}})\right]$, $\mathbb{E}\!\left[C_{(1,1)}(\ln X_{\mathrm{BH}})\right]$ via~(\ref{inttable1})--(\ref{inttable4}), and similarly for the moments over $Y_{\mathrm{BH}}$ in~(\ref{eq:cov}) by replacing $n_1$ by $n_2$ accordingly. This completes the calculation of the desired term $\mathbb{E}\!\left[\operatorname{tr}^2(X_{\mathrm{BH}}\ln Y_{\mathrm{BH}})\right]$ via~(\ref{decom:tr^2xlny}), and consequently the required $\mathbb{V}[\operatorname{tr}(\rho_{\mathrm{BH}} \ln \sigma_{\mathrm{BH}})]$ via~(\ref{eq:tr^2rhosigma}) and~(\ref{eq:Ve2rhosig}). Together with~(\ref{eq:rhomsig}), the derived result on $\mathbb{E}[\operatorname{tr}(\rho_{\mathrm{BH}} \ln \rho_{\mathrm{BH}}) \operatorname{tr}(\rho_{\mathrm{BH}} \ln \sigma_{\mathrm{BH}})]$ via~(\ref{decom:xlnxxlny}), (\ref{decom:xlnxx}) and (\ref{eq:trrholnrhotrrholnsigma}), and known results on $\mathbb{E}\!\left[\operatorname{tr}(\rho_{\mathrm{BH}}\ln \rho_{\mathrm{BH}})\right]$ in~(\ref{EBH}) and $\mathbb{V}\!\left[\operatorname{tr}(\rho_{\mathrm{BH}}\ln \rho_{\mathrm{BH}})\right]$ in~(\ref{VBH}), respectively, we deduce the final result~(\ref{eq:r2}) via~(\ref{eq:rev4}), where we have also used the identity~\cite{Brychkov08}
\begin{eqnarray}
    \psi_0\!\left(x+\frac{1}{2}\right)=2\psi_0(2x)-\psi_0(x)-2\ln 2
\end{eqnarray}
in the simplification of polygamma functions. We then obtain~(\ref{eq:degenr2}) from~(\ref{eq:r2}) by resolving the indeterminacy via~(\ref{eq:pgna}).
\end{proof}

\subsection{Proof of Proposition~\ref{prop:HS-BH} and the other variance results}\label{sec:prhsbh}
\begin{proof}
For variance of relative entropy between BH and HS ensembles, the
corresponding required term $\mathbb{E}\!\left[\operatorname{tr}^2(X_{\mathrm{BH}}\ln Y_{\mathrm{HS}})\right]$ can be read off from the obtained results for $X_{\mathrm{BH}}$ and $Y_{\mathrm{HS}}$ via~(\ref{inttable1})--(\ref{inttable4}) and~(\ref{decom:tr^2xlny}) . The result for $\mathbb{E}[\operatorname{tr}(\rho_{\mathrm{BH}} \ln \rho_{\mathrm{BH}}) \operatorname{tr}(\rho_{\mathrm{BH}} \ln \sigma_{\mathrm{HS}})]$ is also performed as the combination of individual results for $X_{\mathrm{BH}}$ and $Y_{\mathrm{HS}}$ via~(\ref{decom:xlnxxlny}), (\ref{decom:xlnxx}) and (\ref{eq:trrholnrhotrrholnsigma}). Similar to the previous two propositions, we deduce result~(\ref{eq:r3}) via~(\ref{eq:rev4}) and resolve the indeterminacy~(\ref{eq:pgna}) to obtain~(\ref{eq:r3n1n2=m}).
\end{proof}

Similarly, an exact formula of variance of relative entropy of the case $D(\rho_{\rm{HS}}||\sigma_{\rm{BH}})$ can be extracted as
\begin{eqnarray}\label{eq:r4}
\mathbb{V}\!\left[D(\rho_{\rm{HS}}||\sigma_{\rm{BH}})\right]&=&\frac{2}{ m n_1+1}\mathcal{B_{\rm{BH}}} +\mathrm{CF_4},
\end{eqnarray}
where $\mathcal{B_{\rm{BH}}}$ is the finite summation as~(\ref{eq:bbh}) with $\Psi_{a,b}$ as~(\ref{Psi}), and $\mathrm{CF_4}$ is a closed-form expression that we list in Appendix~\ref{appd:closedform}. By resolving indeterminacy~(\ref{eq:pgna}), one also has their degenerate case for $n_1=n_2=m$,
\begin{eqnarray}\label{eq:r4n1n2=m}
&&\mathbb{V}\!\left[D(\rho_{\rm{HS}}||\sigma_{\rm{BH}})\right]\nonumber\\&=&\frac{(m-1) \left(m^3-3 m^2-4 m-4\right)}{4 m^4 \left(m^2+1\right)}+\frac{1}{m^2+1}\left(\frac{3(m-1)}{m} (\psi_0 (m)-\psi_0 (1))+(5 m-2) \psi _1(m)+5 \psi _1(1)\right)\nonumber\\&&-\psi _1\!\left(\frac{m^2}{2}\right)-\psi _1\!\left(m^2\right).
\end{eqnarray}

\subsection{Proof of Corollary~\ref{vanish}}\label{sec:prcoro}
\begin{proof}
The asymptotic behavior of the variances of relative entropy~(\ref{eq:r1}),~(\ref{eq:r2}),~(\ref{eq:r3}) in Corollary~\ref{vanish} is obtained by applying the limiting approximations of polygamma functions to the exact results~(\ref{eq:r1}), (\ref{eq:r2}), (\ref{eq:r3}), respectively,
\begin{eqnarray}
    \psi_0(x)&\sim&\ln x-\frac{1}{2 x}-\sum_{l=1}^{\infty} \frac{B_{2 l}}{2 l x^{2 l}}, \quad x \rightarrow \infty \\
 \psi_1(x)&\sim&\frac{1+2 x}{2 x^2}+\sum_{l=1}^{\infty} \frac{B_{2 l}}{x^{2 l+1}}, \quad x \rightarrow \infty,
\end{eqnarray}
where the constant $B_k$ is the $k$-th Bernoulli number~\cite{Brychkov08}, which yield that assuming $F$~(\ref{eq:assum}) holds,
\begin{eqnarray}
\mathrm{CF}_i\longrightarrow 0, ~~m\to \infty ~~~~i=1,2,3.
\end{eqnarray}
Also, one notices that in the regime~(\ref{eq:limfixn-m}),\begin{equation}
\Psi_{a,b} \longrightarrow \sum_{k=1}^{\infty}\frac{1}{k^2}=\psi_{1}(1)=\frac{\pi^2}{6},~~m\to \infty
\end{equation}
valid for any finite fixed $a$ and $b$~\cite{Wei22}, and similarly assuming $F$~(\ref{eq:assum})
\begin{equation}
\Psi_{0,2 \beta_2+m}\longrightarrow\sum_{k=1}^{\infty}\frac{1}{2^k k^2}=\frac{\pi^2}{12}-\frac{\ln^2 2}{2}, ~~m\to \infty
\end{equation}
which are finite numbers. Hence one clearly has, assuming $F$~(\ref{eq:assum}) holds, the below limiting processes follow,
\begin{eqnarray}
    \mathbb{V}\!\left[D(\rho_{\rm{HS}}||\sigma_{\rm{HS}})\right]&\longrightarrow &\frac{\pi^2}{6}\displaystyle \lim_{m\to \infty}\frac{2}{m n_1+1}=0, ~~m\to \infty\\
        \mathbb{V}\!\left[D(\rho_{\rm{BH}}||\sigma_{\rm{BH}})\right]&\longrightarrow& \left(\frac{\pi^2}{6}-\frac{\pi^2}{6}+\frac{\pi^2}{3}-\frac{\pi^2}{6}+\ln^2 2\right)\displaystyle \lim_{m\to \infty}\frac{m-4 n_1}{n_1 \left(m^2-2 m n_1-2\right)}=0, ~~m\to \infty\\
            \mathbb{V}\!\left[D(\rho_{\rm{BH}}||\sigma_{\rm{HS}})\right]&\longrightarrow& \frac{\pi^2}{6}\displaystyle \lim_{m \to \infty}\frac{m-4 n_1}{n_1 \left(m^2-2 m n_1-2\right)}=0, ~~m\to \infty.
\end{eqnarray}
This completes the proof of the vanishing of variance of relative entropy~(\ref{eq:vanish}).

As a consequence, we can conclude relative entropy~(\ref{eq:red}) as an entropic metric is typical from the vanishing of variance under regime~(\ref{eq:limfixn-m}) if the average relative entropy is already approaching a constant. From the exact results in~\cite{wei2025averagerelativeentropyrandom}, one has obtained that
\begin{eqnarray}
\E\!\left[D(\rho_{\rm{HS}}||\sigma_{\rm{HS}})\right]&=&\psi_{0}\!\left(mn_{2}\right)-\psi_{0}\!\left(mn_{1}+1\right)+\psi_{0}\!\left(n_{1}\right)-\frac{1}{m}\left(n_{2}\psi_{0}\!\left(n_{2}\right)-\alpha_{2}\psi_{0}\!\left(\alpha_{2}\right)\right) \nonumber \\
&&+~\!\frac{m+2n_{1}+1}{2n_{1}}\label{avresult1}\\
\E\!\left[D(\rho_{\rm{BH}}||\sigma_{\rm{BH}})\right]&=&\psi_{0}\!\left(mn_{2}-\frac{m^{2}}{2}\right)-\psi_{0}\!\left(mn_{1}-\frac{m^{2}}{2}+1\right)+\psi_{0}\!\left(n_{1}+\frac{1}{2}\right)\nonumber\\
&&+~\!\frac{n_{2}}{m}\psi_{0}\!\left(n_{2}\right)-\frac{2n_{2}-m}{m}\left(\psi_{0}\!\left(n_{2}-\frac{m}{2}\right)+\psi_{0}\!\left(n_{2}-\frac{m}{2}+\frac{1}{2}\right)\right)\nonumber\\
&&+~\!\frac{\alpha_{2}}{m}\left(\psi_{0}\!\left(\alpha_{2}\right)+2\psi_{0}\!\left(\alpha_{2}+\frac{1}{2}\right)\right)+1\label{avresult2}\\
\E\!\left[D(\rho_{\rm{BH}}||\sigma_{\rm{HS}})\right]&=&\psi_{0}\left(mn_{2}\right)-\psi_{0}\left(mn_{1}-\frac{m^{2}}{2}+1\right)+\psi_{0}\left(n_{1}+\frac{1}{2}\right)\nonumber\\
&&-~\!\frac{1}{m}\left(n_{2}\psi_{0}\left(n_{2}\right)-\alpha_{2}\psi_{0}\left(\alpha_{2}\right)\right)+1.\label{avresult3}
\end{eqnarray}
Therefore, assuming $F$~(\ref{eq:assum}) holds, the below limiting processes follow,
    \begin{eqnarray}\label{eq:HSlim}
\E\!\left[D(\rho_{\rm{HS}}||\sigma_{\rm{HS}})\right]&\longrightarrow&\frac{3}{2} ,~~m\to \infty\\
\E\!\left[D(\rho_{\rm{BH}}||\sigma_{\rm{BH}})\right]&\longrightarrow&1+2\ln 2,~~m\to \infty\\
\E\!\left[D(\rho_{\rm{BH}}||\sigma_{\rm{HS}})\right]&\longrightarrow&1+\ln 2,~~m\to \infty.
% \E\!\left[D(\rho_{\rm{HS}}||\sigma_{\rm{BH}})\right]&\stackrel{(\ref{eq:limfixn-m})}{\longrightarrow}&\frac{3}{2} + \ln 2.
\end{eqnarray}
That is, in the regime~(\ref{eq:limfixn-m}), the limiting average relative entropies are all constants. Together with the vanishing result~(\ref{eq:vanish}), we deduce that relative entropy~(\ref{eq:red}) is typical in the regime~(\ref{eq:limfixn-m}) for HS ensemble~(\ref{eq:HS}), BH ensemble~(\ref{eq:BH}) and between these two ensembles.
\end{proof}

\counterwithin{equation}{section}
\appendices
\section*{Appendices}
\section{Closed-form expressions in Proposition~\ref{prop:HS}--\ref{prop:HS-BH} and (\ref{eq:r4})}\label{appd:closedform}
In this section, we list the closed-form expressions involving polygamma functions in the propositions.
\begin{eqnarray}
     \mathrm{CF_1}  &=& -\frac{m^4+2 m^3 n_1+2 m^3-4 m^2 n_1^2+2 m^2 n_1+m^2-4 m n_1-4}{4 m^2 n_1^2 (m n_1+1)}\nonumber\\ &&+\frac{1}{mn_1+1}\bigg(\frac{2 m+n_1 n_2 }{m n_2}\left(\psi_0 (n_2)-\psi_0(n_2-m)\right)+\frac{m^2-n_2^2}{m^2}\left(\psi_0(n_2)-\psi_0(n_2-m)\right){}^2\nonumber\\ &&+\frac{n_1(m-n_2) }{m}\psi_1(n_2-m)+(m+n_1) \psi_1(n_1)+\frac{m+n_1 n_2}{m} \psi_1(n_2)\!\bigg)\!-\psi_1(m n_1)-\psi_1 (m n_2)\\
  \mathrm{CF_2}  &=& \frac{m^5-8 m^4 n_1+20 m^3 n_1^2-16 m^2 n_1^3+8 m^2 n_1-16 m n_1^2-16 n_1}{2 m^2 n_1 \left(m-2 n_1\right){}^2 \left(m^2-2 m n_1-2\right)}\nonumber\\ &&
  + \frac{1}{n_1(m^2-2mn_1-2)}
\bigg(\frac{(m-2n_1+1)(2n_1-1)}{2m}(\psi_0(n_2-m)-\psi_0(n_2)\nonumber\\ &&+4(\psi_0(2n_2-m)-\psi_0(2n_2-2m)))
+\frac{m-4n_1}{4}(8\psi_0(2n_2-m)\psi_0(2n_2)+\psi_0^2(n_2)-4\psi_0^2(2n_2)\nonumber\\ &&-4\psi_1(2n_2)+6\psi_1(1))
-\frac{n_2(m-4n_1)(m-n_2)}{2m^2}((-\psi_0(n_2-m)+4\psi_0(2n_2-2m)+2\psi_0(n_2))\nonumber\\ &&\times(\psi_0(n_2-m)-4\psi_0(2n_2-2m))+8(-\psi_0(n_2-m)+4\psi_0(2n_2-2m)+\psi_0(n_2))\psi_0(2n_2-m))
\nonumber\\ &&-\frac{(m-4n_1)(m^2-2mn_2+2n_2^2)}{4m^2}\psi_0^2(n_2)-\frac{(m-4n_1)(m^2-8mn_2+8n_2^2)}{m^2}\psi_0^2(2n_2-m)
\nonumber\\ &&-\frac{4 m^2 n_1-m^2-8 m n_1^2+4 m n_1-8 m n_1 n_2-2 m+16 n_1^2 n_2+4 n_2}{m}\psi_1(2n_2-m)
\nonumber\\ &&-\frac{m^2-2mn_1-4n_1^2-1}{2}(\psi_1(n_1)-4\psi_1(2n_1)) -\frac{n_2(2mn_1-4n_1^2-1)}{2m}\psi_1(n_2)
\nonumber\\ &&-\frac{(2mn_1-4n_1^2-1)(m-n_2)}{2m}\left(\psi_1(n_2-m)-8\psi_1(2n_2-2m)\right)
\bigg)
\nonumber\\ &&-\psi_1\left(mn_1-\frac{m^2}{2}\right)-\psi_1\left(mn_2-\frac{m^2}{2}\right)\\
   \mathrm{CF_3}  &=&  \frac{m^5-8 m^4 n_1+20 m^3 n_1^2-16 m^2 n_1^3+8 m^2 n_1-16 m n_1^2-16 n_1}{2 m^2 n_1 \left(m-2 n_1\right){}^2 \left(m^2-2 m n_1-2\right)}\nonumber\\ &&+\frac{1}{n_1 \left(m^2-2 m n_1-2\right)}\bigg(\frac{2 m^2-8 m n_1+2 m n_1 n_2-4 n_1^2 n_2-n_2 }{2 m n_2}\left( \psi_0(n_2)-\psi_0 (n_2-m) \right)\nonumber\\ &&+\frac{\left(m-4 n_1\right) \left(m^2-n_2^2\right) }{2m^2}\left( \psi_0 (n_2)- \psi_0 (n_2-m)\right)^2\nonumber\\ &&+\frac{\left(2 m n_1-4 n_1^2-1\right) \left(m-n_2\right) }{2 m}\psi _1(n_2-m)+\frac{m^2-4 m n_1+2 m n_1 n_2-4 n_1^2 n_2-n_2 }{2 m}\psi _1(n_2)\nonumber\\ &&-\left(m^2-2 m n_1-4 n_1^2-1\right) \left(\frac{1}{2} \psi _1(n_1)-2 \psi _1(2 n_1)\right)\bigg)-\psi _1\!\left(m n_1-\frac{m^2}{2}\right)-\psi _1(m n_2)
   \end{eqnarray}
\begin{eqnarray}
 \mathrm{CF_4}  &=&    -\frac{m^4+2 m^3 n_1+2 m^3-4 m^2 n_1^2+2 m^2 n_1+m^2-4 m n_1-4}{4 m^2 n_1^2 (m n_1+1)}\nonumber\\&&
  +\frac{1}{m n_1+1}\bigg(-\frac{n_1-1 }{m}(\psi _0(n_2)-\psi _0(n_2-m)+4 (\psi _0(2 n_2-2 m)-\psi _0(2 n_2-m)))\nonumber\\&&-\frac{2 (m^2-8 m n_2+8 n_2^2)}{m^2} \psi _0^2(2 n_2-m)+\frac{ n_2 (m-n_2)}{m^2}(-2 \psi _0(n_2) (\psi _0(n_2-m)+4 \psi _0(2 n_2-m)) \nonumber\\&& + \psi _0^2(n_2-m)+\psi _0^2(n_2)+ 8(\psi _0(2 n_2-m) (\psi _0(n_2-m)-4 \psi _0(2 n_2-2 m))\nonumber\\&&+\psi _0(2 n_2-2 m) (-\psi _0(n_2-m)+2 \psi _0(2 n_2-2 m)+\psi _0(n_2))))+4 \psi _0(2 n_2) \psi _0(2 n_2-m)\nonumber\\&&-2 \psi _0^2(2 n_2) -\frac{1}{m}(2 (2 m n_1-m-4 n_1 n_2) \psi _1(2 n_2-m)+n_1 (m-n_2) (\psi _1(n_2-m)\nonumber\\&&-8 \psi _1(2 n_2-2 m))+n_1 n_2 \psi _1(n_2))+3\psi _1(1)-2\psi _1(2n_2)+(m+n_1)\psi _1(n_1)\bigg)
  \nonumber\\&&-\psi _1\left(m n_2-\frac{m^2}{2}\right)-\psi _1\left(m n_1\right)
\end{eqnarray}

\section{Finite summation of integrals~(\ref{eq:IA})--(\ref{eq:ID})}\label{appd:finsum}
In this appendix, we present the finite summations of integrals from Cauchy-Laguerre two-point kernels with Pochhammer's symbol~(\ref{eq:poch}), and the removable singularities are interpreted by limits $\epsilon\to0$.

\begin{equation}\label{eq:IA-pochhammer}
\begin{aligned}
\mathrm{I}_{\mathrm A}(t)
={}&-\sum_{j=0}^{m-1}
\frac{(-1)^{j+m}(2t+j-m+1)_m
      (2\beta+j+2)_m(\beta+j+1)_{2t}}
     {2t\,j!\,(m-j-1)!\,(2\beta+2t+j+2)_m}
\\
&-\sum_{j=0}^{m-1}
\frac{(-1)^{j+m}(2t+j-m+1)_m
      (2\beta+j+2)_m(\beta+j+2)_{2t}}
     {2t\,j!\,(m-j-1)!\,(2\beta+2t+j+2)_m}.
\end{aligned}
\tag{B.1}
\end{equation}

\begin{equation}\label{eq:IB-pochhammer}
\begin{aligned}
\mathrm{I}_{\mathrm B}(t)
={}&\lim_{\epsilon\to0}
\sum_{i=0}^{m-1}\sum_{j=0}^{m-1}
\frac{(-1)^{i+j}
      (t+i-m+\epsilon+1)_{m-\epsilon}
      (t+j-m+\epsilon+1)_{m-\epsilon}}
     {i!\,j!\,(m-i-1)!\,(m-j-1)!}
\\
&\quad\times
\frac{(2\beta+i+2)_m(2\beta+j+2)_m
      (\beta+i+1)_t(\beta+j+1)_t}
     {(2\beta+t+i+2)_m(2\beta+t+j+2)_m}
\\
&\quad\times
\frac{1}{(-t+i-j-\epsilon)(-t-i+j-\epsilon)}.
\end{aligned}
\tag{B.2}
\end{equation}

\begin{equation}\label{eq:IC-pochhammer}
\begin{aligned}
\mathrm{I}_{\mathrm C}(t)
={}&\lim_{\epsilon\to0}
\sum_{i=0}^{m-1}\sum_{j=0}^{m-1}
\frac{(-1)^{i+j}
      (t+i-m+\epsilon+1)_{m-\epsilon}
      (t+j-m+\epsilon+1)_{m-\epsilon}}
     {i!\,j!\,(m-i-1)!\,(m-j-1)!}
\\
&\quad\times
\frac{(2\beta+i+2)_m(2\beta+j+2)_m
      (\beta+i+2)_t(\beta+j+2)_t}
     {(2\beta+t+i+2)_m(2\beta+t+j+2)_m}
\\
&\quad\times
\frac{1}{(-t+i-j-\epsilon)(-t-i+j-\epsilon)}.
\end{aligned}
\tag{B.3}
\end{equation}

\begin{equation}\label{eq:ID-pochhammer}
\begin{aligned}
\mathrm{I}_{\mathrm D}(t)
={}&4\sum_{j=0}^{m-1}\sum_{k=0}^{m-1}
       \sum_{i=0}^{j}\sum_{s=0}^{j}
\frac{(-1)^{i+s}(\beta+j+1)(\beta+k+1)}
     {i!\,s!\,(j-i)!\,(j-s)!}
\\
&\quad\times
\frac{(2\beta+i+2)_j(2\beta+s+2)_j}
     {(2\beta+t+i+2)_{k+1}(2\beta+t+s+2)_{k+1}}
\\
&\quad\times
(\beta+i+1)_t(\beta+s+2)_t
(t+i-k+1)_k(t+s-k+1)_k
\\
&-2\sum_{j=0}^{m-1}\sum_{i=0}^{j}\sum_{k=0}^{j}
\frac{(-1)^{i+k}(\beta+j+1)}
     {i!\,k!\,(j-i)!\,(j-k)!\,(2\beta+2t+i+k+2)}
\\
&\quad\times
(2\beta+i+2)_j(2\beta+k+2)_j
(\beta+i+1)_t(\beta+k+2)_t.
\end{aligned}
\tag{B.4}
\end{equation}
% Experimental section

% \section{Experimental Section}
% \threesubsection{First part of experimental section}\\
% \threesubsection{Second part of experimental section}\\

\medskip
% Acknowledgements
\medskip
\textbf{Acknowledgements} \par %delete if not applicable))
The work of Lu Wei was supported by the U.S. National Science Foundation (2306968) and the U.S. Department of Energy (DE-SC0024631).

% References
\medskip

\end{document}